\documentclass{article}
\usepackage{arxiv}
\usepackage[T1]{fontenc}
\usepackage{amsmath,amssymb,booktabs,microtype,array}
\usepackage{float,placeins,graphicx,pdflscape}
\usepackage{tikz}
\usetikzlibrary{arrows.meta,calc}
\tikzset{
  paper diagram/.style={font=\normalfont\rmfamily\small},
  paper diagram note/.style={font=\normalfont\rmfamily\footnotesize},
  paper diagram heading/.style={font=\normalfont\rmfamily\small\bfseries}
}
\usepackage[hidelinks]{hyperref}
\graphicspath{{docs/figures/physical_edmd/}{figures/physical_edmd/}}
\newcommand{\R}{\mathbb R}
\newcommand{\E}{\mathbb E}
\makeatletter
\def\input@path{{figures/physical_edmd/}{docs/figures/physical_edmd/}}
\makeatother
\title{Daily Wildfire Prediction with Spatial EDMD}
\author{
  Ion Matei\\
  Fujitsu Research of America\\
  \texttt{imatei@fujitsu.com}
  \And
  Maksym Zhenirovskyy\\
  Fujitsu Research of America\\
  \texttt{mzhenirovskyy@fujitsu.com}
  \And
  Hon Yung Wong\\
  Fujitsu Research of America\\
  \texttt{awong@fujitsu.com}
}
\date{}
\renewcommand{\undertitle}{}
\renewcommand{\headeright}{}
\hypersetup{
  pdftitle={Daily Wildfire Prediction with Spatial EDMD},
  pdfauthor={Ion Matei, Maksym Zhenirovskyy, Hon Yung Wong}
}
\begin{document}
\maketitle
\begin{abstract}
We forecast wildfire spread from daily satellite-based fire-progression maps via extended dynamic mode decomposition (EDMD), an approximation of the stochastic Koopman operator. This operator maps functions of the
state, called observables, to their expected values one day later. First, we compute seven
observables at every cell: the daily growth and six are exposures (i.e., averages of the daily growth over a neighborhood of an unburned cell). The EDMD linear operator, fit  by ridge regression, advances the observable fields by one day with 22 fixed
spatial filters and matrices shared by all cells.  Second,
we construct two predictors that convert the advanced observables into a
daily ignition probability for every unburned cell. Each cell carries
the probabilities of being unburned, newly burned, or previously burned.
Both predictors keep these probabilities nonnegative with unit sum, keep
burned cells burned, and prevent ignition where the fuel map is
nonburnable. The hybrid predictor feeds the resulting growth back into the
operator. The learned decoder predictor advances the observables linearly and maps
them to ignition probabilities. Third, we
fit both predictors to the first 11 maps of six fires, for each fire
separately and jointly over all six (pooled). From the eleventh map, we
forecast each fire to the end of its record, 5 to 27 days ahead, without
further observations. We compare with persistence, a forecast of no
further growth, and with CA--CNN, a cellular automaton whose spread
parameters a convolutional neural network predicts from terrain, fuel,
and wind~\cite{cacnn}. On cells unburned at the forecast start, the burn
probability of the pooled learned decoder has a mean squared error of
0.0703, against 0.1123 for persistence and 0.1262 for pooled CA--CNN.
After thresholding at a level chosen on the first two forecast days, the
decoder attains an intersection over union with the observed burned area
of 0.374, against 0.393 for pooled CA--CNN. The EDMD predictors thus reach comparable
accuracy at a fraction of the CA--CNN memory and training time. The
largest EDMD model has 1,085 coefficients against 36,018 CA--CNN
parameters, and trains in under 30\,s on one GPU against 10 to 15\,h for
CA--CNN.
\end{abstract}

\section{Introduction}
\label{sec:intro}
Our objective is to forecast the progression of wildfires using daily satellite observations. These observations provide a record of a wildfire's
progression as daily burned-area maps. We call each map pixel a cell,
about 30\,m on a side. We design predictors that learn from this record, forecast
subsequent burned area, and can be refitted as observations arrive. Their
forecasts must preserve burned cells and prevent ignition on nonburnable
fuel. We study daily burned-area maps of six fires from our earlier study ~\cite{cacnn}, where we used a cellular automaton (CA) combined with a convolutional neural network (CNN) to predict spread parameters from terrain, fuel, and wind. We compare our new predictors with the CA--CNN, and with a persistence baseline.

We use the stochastic Koopman operator to describe the uncertain daily
evolution~\cite{koopman1931,mezic2005,brunton2022}. It maps functions of
the state, called observables, to their conditional expectations one day
ahead. Extended dynamic mode decomposition (EDMD) estimates a
finite-dimensional approximation by least squares~\cite{williams}.
We propagate observables of the two latest maps and convert their
predictions into admissible cell-state probabilities.

This work makes three contributions. First, we construct seven spatial
observables: recent growth and six exposures that measure nearby growth
at unburned cells. We parameterize their propagation with 22 spatial
filters and matrices shared across cells. The operator has at most 1,078
coefficients, independent of grid size. Second, we construct two
predictors with nonnegative cell-state probabilities, irreversible
transitions to burned states, and zero ignition on nonburnable fuel. The
hybrid predictor feeds accepted growth back into the operator. The learned decoder predictor
propagates the observables linearly and maps them to ignition probabilities
with seven fitted coefficients. Third, we fit each fire separately
(per-event) and all six jointly (pooled). We evaluate continuous forecasts,
daily restarts, and expanding-history refits. Persistence provides a
no-growth baseline, while the
CA--CNN provides the learned reference~\cite{cacnn}.

Both predictors improve aggregate probability scores over the supplied
CA--CNN checkpoints. The hybrid achieves the highest per-event
intersection over union (IoU), while pooled CA--CNN retains the highest
pooled IoU. The compact EDMD models fit in seconds, which supports
refitting after each new observation.

\paragraph{Paper roadmap.}
Section~\ref{sec:problem} defines the data and prediction task.
Sections~\ref{sec:koopman}, \ref{sec:physical}, and~\ref{sec:learning}
formulate the operator, predictors, and estimation.
Sections~\ref{sec:protocol} and~\ref{sec:results} present the evaluation
and results. Sections~\ref{sec:discussion} and~\ref{sec:conclusion}
discuss related work, implications, and conclusions.
Appendix~\ref{app:dynamics} analyzes the learned decoder's linear
dynamics, and Appendix~\ref{app:refitting} details daily restarts and refits.

\paragraph{Notation.}
We write $\R$ for the real numbers, $\E$ for expectation, and $I$ for the
identity map. Superscripts $\top$, $*$, and $\dagger$ denote transpose,
conjugate transpose, and Moore--Penrose pseudoinverse. The norms
$\|\cdot\|_2$ and $\|\cdot\|_F$ are the Euclidean or spectral norm and the
Frobenius norm, respectively. The symbol $\odot$, divisions, exponentials, maxima,
and inequalities between fields act cellwise. We define
$[z]_+=\max(0,z)$ and $\operatorname{clip}(z,0,1)=\min(1,[z]_+)$.
The notation $|S|$ gives the cardinality of the set $S$.
For matrix $A$, $\operatorname{tr}(A)$ denotes its trace and $\otimes$ represents the Kronecker product.

\section{Prediction problem and data}
\label{sec:problem}
This section defines the daily maps and cell states, describes the six
fires and their data, and states the forecast task and the training dates.

Let $\Omega$ be the rectangular grid of a fire and let $N=|\Omega|$.
We index the $n_{\rm obs}$ observed maps by $i=1,\ldots,n_{\rm obs}$ in time
order, and we denote by $d_i$ the date of map $i$. We count calendar days from
day 1, the date $d_1$ of the first map. We reserve
$t$ for calendar time. Each model step advances one calendar day.
The binary map $c_t(x)$ indicates whether cell $x\in\Omega$ belongs to the
cumulative burned area on day $t$. The maps are nondecreasing in time. We define the observed daily growth
and the two-map history, whenever both daily maps exist, by
\begin{equation*}
 g_t=c_t-c_{t-1},\qquad X_t=(c_t,c_{t-1}).
\end{equation*}
Monotonicity and binary values imply
\begin{equation}
 g_t=(1-c_{t-1})\odot c_t.
 \label{eq:growth-identity}
\end{equation}
We use successive maps to define three daily states:
\begin{equation}
 u_t=1-c_t,\qquad b_t=g_t,\qquad r_t=c_{t-1}.
 \label{eq:observed-state}
\end{equation}
Here $u_t$ is unburned, $b_t$ is newly burned during the preceding day,
and $r_t$ is previously burned. A forecast
represents these quantities by nonnegative fractions or marginal
probabilities with $u_t+b_t+r_t=1$. Thus each cell lies on the
probability simplex. We use the observed identities above to construct
training targets and initialize forecasts. 

We study six fire events: Bear (2020)~\cite{bear2020}, Brattain
(2020)~\cite{brattain2020}, Buck (2017)~\cite{buck2017}, Chimney
(2016)~\cite{chimney2016}, Ferguson (2018)~\cite{ferguson2018}, and Pier
(2017)~\cite{pier2017}. We use the daily fire-progression maps
and aligned environmental rasters distributed with the reference
CA--CNN study~\cite{cacnn,pytorchfire}. The reference study derives the fire maps from MODIS observations and
represents them on a computational grid of about 30\,m. The MODIS measurements are much coarser than this grid, so the maps do
not resolve features at the scale of one cell. LANDFIRE fuel classes identify nonburnable
cells~\cite{landfire,scott2005} and define the fuel mask of
Section~\ref{sec:physical}. EDMD receives the map history and a fixed
fuel mask. CA--CNN additionally receives terrain and recorded daily wind.



Let $t_0$ be the date of the latest observed map supplied at a forecast
start. Both predictors initialize from $X_{t_0}$ and predict
$c_{t_0+h}$ for integer horizons $h\ge1$. Continuous forecasts start at the training end
$t_0=d_{11}$ (i.e., the eleventh recorded map). Buck has no maps on days 8 and 9, so its $d_{11}$ is day 13.
Let $\mathcal D=\{d_i:1\le i\le n_{\rm obs}\}$ be the observed dates.
For a training end $d_i$, define the eligible daily training targets
and the available forecast horizons by
\begin{align*}
 \mathcal P_i&=\{t\in\mathcal D:t\le d_i,\ t-1,t-2\in\mathcal D\},\\
 \mathcal H_i^{(H)}&=\{h\in\{1,\ldots,H\}:d_i+h\in\mathcal D\}.
\end{align*}
Here $H$ is the maximum forecast horizon. A target $t\in\mathcal P_i$
supplies the pair $(X_{t-1},X_t)$. A restart at $d_i$ requires
$d_i-1\in\mathcal D$. We skip starts without that preceding daily map.
We propagate across missing target dates and score only the observed
horizons in $\mathcal H_i^{(H)}$. 

\begin{table}[H]\centering\small
\caption{Data and temporal split. Training includes maps through $d_{11}$.
Daily EDMD pairs require three consecutive maps. Test-date counts follow
the calibration split in Section~\ref{sec:protocol}.}
\label{tab:data}
\begin{tabular}{lrrlr}\toprule
Fire & Grid & Daily pairs & Date $d_{11}$ & Test dates\\\midrule
Bear & $287\times415$ & 9 & 2020-08-29 & 10\\
Brattain & $752\times676$ & 9 & 2020-09-17 & 9\\
Buck & $455\times473$ & 7 & 2017-09-25 & 17\\
Chimney & $557\times653$ & 9 & 2016-08-23 & 3\\
Ferguson & $983\times822$ & 9 & 2018-07-23 & 17\\
Pier & $679\times706$ & 9 & 2017-09-08 & 18\\
\bottomrule\end{tabular}
\end{table}

\section{Koopman approximation and spatial observables}
\label{sec:koopman}
This section introduces the stochastic Koopman operator and its EDMD
approximation, the seven observables, and the 22 spatial filters of the
estimated operator.

\subsection{Finite-dimensional Koopman approximation}
We assume that the two-map history $X_t$ follows a time-homogeneous
Markov law. The stochastic Koopman operator maps an integrable observable
$f$ to its conditional expectation one day ahead~\cite{mezic2005,brunton2022}:
\begin{equation*}
 (\mathcal U f)(X)=\E[f(X_{t+1})\mid X_t=X].
\end{equation*}
It is linear in $f$ even when the state dynamics are nonlinear. Applied
to a binary indicator, it returns a probability.

Let $\Phi=(\phi_1,\ldots,\phi_d)^\top$ collect $d$ scalar observables.
For $q$ observed pairs $(X_i,X_i^+)$ one day apart, define
\begin{equation*}
 G=\frac1q\sum_{i=1}^q\Phi(X_i)\Phi(X_i)^\top,\qquad
 C=\frac1q\sum_{i=1}^q\Phi(X_i^+)\Phi(X_i)^\top.
\end{equation*}
Unregularized EDMD fits $K=CG^\dagger$ by least squares~\cite{williams}.
Let $\widehat\Phi_t$ denote the predicted observables. From an observed
initial history $X_{t_0}$, EDMD propagates them by
\begin{equation*}
 \widehat\Phi_{t_0}=\Phi(X_{t_0}),\qquad
 \widehat\Phi_{t+1}=K\widehat\Phi_t,\qquad
 \widehat\Phi_{t_0+h}=K^h\Phi(X_{t_0}).
\end{equation*}
If the span of the observables is invariant under $\mathcal U$, let $K_*$
be its exact matrix on that span. The Markov assumption then gives
$K_*^h\Phi(X_{t_0})=\E[\Phi(X_{t_0+h})\mid X_{t_0}]$.
The fitted $K$ approximates this matrix. Under suitable sampling
assumptions, EDMD converges to the orthogonal projection of $\mathcal U$
onto the observable span as the sample grows~\cite{korda2018}.
Our spatial regression further restricts the operator's coefficients and
coupling. We assess this approximation through multiday forecasts.

\subsection{Neighborhood observables}
\label{sec:lift}
For an odd integer $w$, let $Q_w(x)$ be the $w\times w$ square centered at
$x$. For a scalar grid field $\zeta:\Omega\to\R$, we define the zero-padded
box average by
\begin{equation*}
 [A_w\zeta](x)=\frac1{w^2}\sum_{y\in Q_w(x)\cap\Omega}\zeta(y).
\end{equation*}
We use seven spatial filters $T_0,\ldots,T_6$. The identity is $T_0=I$,
and the box averages are $T_1=A_3$ and $T_6=A_{63}$. Let $\xi_2$ and
$\xi_4$ be the unit grid steps to the north and south, and $\xi_3$ and
$\xi_5$ those to the east and west. For $k=2,\ldots,5$, we set
$[T_k\zeta](x)=[A_{15}\zeta](x+8\xi_k)$ when the shifted center is in
$\Omega$, and zero otherwise. For $k=1,\ldots,6$, we define the exposure
$e_{k,t}$ and stack the seven observables into $v_t\in\R^{7\times N}$:
\begin{equation}
 e_{k,t}=(1-c_t)\odot T_k g_t,\qquad
 v_t=(g_t,e_{1,t},\ldots,e_{6,t})^\top.
 \label{eq:lift}
\end{equation}

An exposure measures recent growth near an unburned cell. The lifting map
$\mathcal V(X_t)=v_t$ contains $7N$ scalar observables and is nonlinear
because of the products in Equation~\eqref{eq:lift}. We compute it from
observed maps for training and forecast initialization. During a forecast,
we propagate its coordinates as \textit{latent observables}, which can
differ from the lift evaluated on the predicted maps.

\subsection{Spatial basis for the propagation operator}
\label{sec:filters}
\begin{figure}[H]\centering
\input{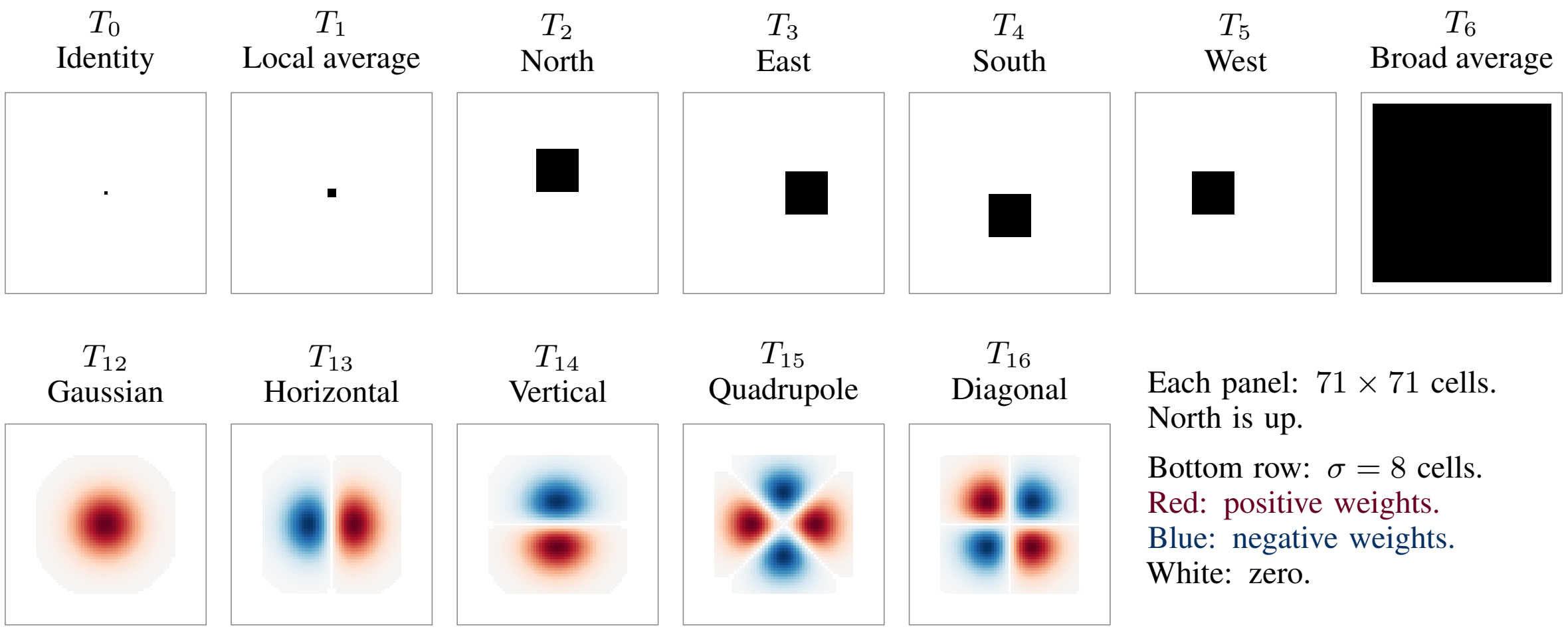}
\caption{Prescribed propagation kernels. The top row shows the identity
and six box filters. The bottom row shows the five Gaussian components at
scale 8 cells. Red denotes positive and blue negative weights. Each kernel
has its own color scale.}
\label{fig:filters}
\end{figure}

We use $J=22$ filters to parameterize the spatial coupling
(Figure~\ref{fig:filters}). Each filter replaces the value at a cell by a
weighted sum of the values at nearby cells. We call these weights, indexed
by the offset from the cell, the kernel of the filter. Alongside $T_0,\ldots,T_6$, the filters
$T_7,\ldots,T_{21}$ contain five Gaussian components at each scale
$\sigma\in\{2,8,16\}$ cells. The Gaussian average measures smooth spatial
variation. Its horizontal and vertical derivatives represent directional
differences. The quadrupole contrasts horizontal and vertical variation,
and the mixed derivative contrasts diagonal orientations. These components
represent local, intermediate, and broad spatial structure with few
coefficients. We truncate each kernel at the smaller of three Gaussian
widths and 31 cells in each coordinate and use zero padding. We give their
formulas in Appendix~\ref{app:mechanics}.

Each $T_k$ acts separately on the seven fields. A matrix
$W_k\in\R^{7\times7}$ mixes the filtered fields, giving the spatial operator
\begin{equation}
 [Lv](x)=\sum_{k=0}^{J-1}W_k[T_kv](x).
 \label{eq:L}
\end{equation}
The matrices are shared across cells, so $49J=1{,}078$ coefficients
specify $L$ on any grid. At each cell, the 22 filtered seven-component
vectors supply 154 regression predictors. Each filtered value is a linear
combination of neighboring observable coordinates. Thus the 22 filters
parameterize the coupling among the fixed $7N$ observables.

We fit $L$ by ridge regression (Section~\ref{sec:learning}) to approximate
$\E[\mathcal V(X_{t+1})\mid X_t]\approx L\mathcal V(X_t)$.
This structured counterpart of $K$ can predict coordinates outside
$[0,1]$. The following predictors convert them into admissible ignition
probabilities.

\section{Physically admissible predictors}
\label{sec:physical}
This section converts the propagated observables into cell-state
probabilities. Section~\ref{sec:transition} defines the transition that
both predictors share, Section~\ref{sec:hybrid} the hybrid predictor, and
Section~\ref{sec:decoder} the learned decoder.

\subsection{A common admissible transition}
\label{sec:transition}
Both predictors initialize from the observed state in
Equation~\eqref{eq:observed-state} and the lift $\mathcal V(X_{t_0})$.
For $t>t_0$, the symbols $u_t,b_t,r_t,c_t$ denote forecasts.
Let $m(x)\in\{0,1\}$ be the fixed burnability mask. We set $m=0$ for
LANDFIRE classes NB1, NB2, NB3, NB8, and NB9, with codes 91, 92, 93, 98,
and 99~\cite{scott2005}. Code 98 denotes open water. The mask restricts
new ignition and preserves the observed initial map where it conflicts
with the fuel classification.

Let $p_t(x)\in[0,1]$ denote the probability of first entering the burned
area during day $t$ to $t+1$, conditional on remaining unburned at $t$.
Each predictor computes $p_t$ from the propagated observables and sets it
to zero where $m=0$. Both then apply
\begin{equation}
 u_{t+1}=(1-p_t)\odot u_t,\qquad
 b_{t+1}=p_t\odot u_t,\qquad
 r_{t+1}=r_t+b_t,\qquad c_{t+1}=b_{t+1}+r_{t+1}.
 \label{eq:transition}
\end{equation}
We call $b_{t+1}$ the \textit{accepted growth}. The corresponding
cell-state transition is a column-stochastic matrix:
\begin{equation*}
 \begin{pmatrix}u_{t+1}\\b_{t+1}\\r_{t+1}\end{pmatrix}
 =\begin{pmatrix}1-p_t&0&0\\p_t&0&0\\0&1&1\end{pmatrix}
 \begin{pmatrix}u_t\\b_t\\r_t\end{pmatrix}
\end{equation*}
that guarantees that the cell state stays on the probability simplex.
Since $c_{t+1}-c_t=p_t\odot u_t\ge0$, we obtain
$c_{t_0}\le c_t\le c_{t+1}\le1$. Previously burned cells remain burned,
and $m=0$ prevents new ignition. These guarantees hold for any fitted
operator coefficients. The forecast cumulative burned area is
\begin{equation*}
 c_{t_0+h}=c_{t_0}+\sum_{j=1}^{h}b_{t_0+j}.
\end{equation*}

\subsection{Hybrid predictor: feedback of accepted growth}
\label{sec:hybrid}
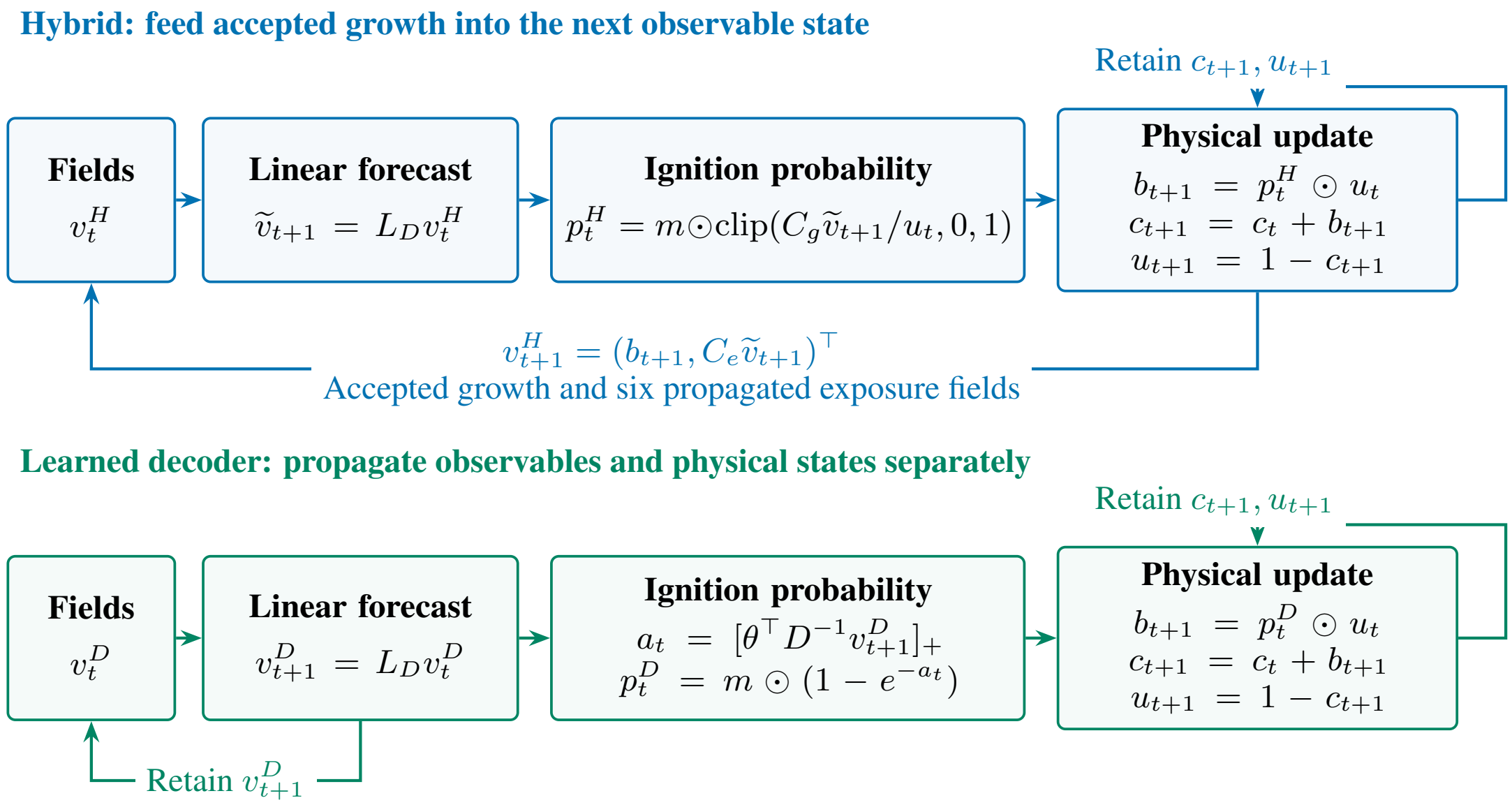
\begin{figure}[H]\centering
\begin{tikzpicture}[
  paper diagram,
  box/.style={draw,rounded corners=2pt,line width=.7pt,
    align=center,inner sep=4pt,minimum height=1.55cm},
  flow/.style={-{Stealth[length=2mm]},line width=.8pt},
  note/.style={fill=white,inner sep=3pt,align=center,paper diagram note}
]
\definecolor{hybridblue}{HTML}{0072B2}
\definecolor{decodergreen}{HTML}{008563}

\begin{scope}[draw=hybridblue]
\node[anchor=west,text=hybridblue,paper diagram heading] at (0,1.65)
  {Hybrid: feed accepted growth into the next observable state};
\node[box,text width=1.3cm,fill=hybridblue!4] (hv) at (.8,0)
  {\textbf{Fields}\\[5pt]$v_t^H$};
\node[box,text width=2.7cm,fill=hybridblue!4] (hl) at (3.35,0)
  {\textbf{Linear forecast}\\[5pt]$\widetilde v_{t+1}=L_Dv_t^H$};
\node[box,text width=4.2cm,fill=hybridblue!4] (hp) at (7.4,0)
  {\textbf{Ignition probability}\\[5pt]
   $p_t^H=m\odot\operatorname{clip}(C_g\widetilde v_{t+1}/u_t,0,1)$};
\node[box,text width=3.5cm,fill=hybridblue!4] (hc) at (11.85,0)
  {\textbf{Physical update}\\[3pt]
   $b_{t+1}=p_t^H\odot u_t$\\
   $c_{t+1}=c_t+b_{t+1}$\\
   $u_{t+1}=1-c_{t+1}$};
\draw[flow] (hv) -- (hl);
\draw[flow] (hl) -- (hp);
\draw[flow] (hp) -- (hc);
\draw[flow] (hc.south) -- (11.85,-1.6) -- (.8,-1.6) -- (hv.south);
\node[note,text=hybridblue] at (6.3,-1.6)
  {$v_{t+1}^H=(b_{t+1},C_e\widetilde v_{t+1})^\top$\\
   Accepted growth and six propagated exposure fields};
\draw[flow] (hc.east) -- ++(.45,0) |- ($(hc.north)+(0,.2)$) -- (hc.north);
\node[note,anchor=west,text=hybridblue] at (10.2,1.3)
  {Retain $c_{t+1},u_{t+1}$};
\end{scope}

\begin{scope}[yshift=-4.15cm,draw=decodergreen]
\node[anchor=west,text=decodergreen,paper diagram heading] at (0,1.65)
  {Learned decoder: propagate observables and physical states separately};
\node[box,text width=1.3cm,fill=decodergreen!4] (dv) at (.8,0)
  {\textbf{Fields}\\[5pt]$v_t^D$};
\node[box,text width=2.7cm,fill=decodergreen!4] (dl) at (3.35,0)
  {\textbf{Linear forecast}\\[5pt]$v_{t+1}^D=L_Dv_t^D$};
\node[box,text width=4.2cm,fill=decodergreen!4] (dp) at (7.4,0)
  {\textbf{Ignition probability}\\[3pt]
   $a_t=[\theta^\top D^{-1}v_{t+1}^D]_+$\\
   $p_t^D=m\odot(1-e^{-a_t})$};
\node[box,text width=3.5cm,fill=decodergreen!4] (dc) at (11.85,0)
  {\textbf{Physical update}\\[3pt]
   $b_{t+1}=p_t^D\odot u_t$\\
   $c_{t+1}=c_t+b_{t+1}$\\
   $u_{t+1}=1-c_{t+1}$};
\draw[flow] (dv) -- (dl);
\draw[flow] (dl) -- (dp);
\draw[flow] (dp) -- (dc);
\draw[flow] (dl.south) -- (3.35,-1.35) -- (.8,-1.35) -- (dv.south);
\node[note,text=decodergreen] at (2.075,-1.35) {Retain $v_{t+1}^D$};
\draw[flow] (dc.east) -- ++(.45,0) |- ($(dc.north)+(0,.2)$) -- (dc.north);
\node[note,anchor=west,text=decodergreen] at (10.2,1.3)
  {Retain $c_{t+1},u_{t+1}$};
\end{scope}
\end{tikzpicture}
\caption{Daily forecast updates. The hybrid feeds accepted growth and
propagated exposures into the next operator step. The learned decoder
propagates all seven latent observables linearly. Both update the cell
states through an ignition probability and start from the same observed
map pair.}
\label{fig:forecast-recurrence}
\end{figure}

Let $L_D$ be the operator of Equation~\eqref{eq:L} fitted to all seven
observables, used by both predictors. Let $C_gv$ select the growth
component of a seven-component vector $v$, and let $C_ev$ select its six
exposures. The hybrid converts predicted growth into an ignition
probability and feeds accepted growth into the next step
(Figure~\ref{fig:forecast-recurrence}).

By Equation~\eqref{eq:growth-identity}, expected next-day growth satisfies
\begin{equation*}
 \E[g_{t+1}(x)\mid X_t]=(1-c_t(x))\,\Pr\big(c_{t+1}(x)=1\mid X_t\big).
\end{equation*}
On an unburned cell, the second factor is its ignition probability.
Regression fits the left side. During forecasting, the hybrid uses the
current unburned fraction $u_t$ to convert predicted growth into an
ignition probability. Starting from $v_{t_0}^H=\mathcal V(X_{t_0})$, it
computes $\widetilde v_{t+1}=L_Dv_t^H$ and applies
\begin{equation*}
 p_t^H=m\odot\operatorname{clip}\big(C_g\widetilde v_{t+1}/u_t,0,1\big),\qquad
 v_{t+1}^H=(b_{t+1},C_e\widetilde v_{t+1})^\top.
\end{equation*}
We set $p_t^H=0$ where $u_t=0$. Since ridge regression does not constrain
the predicted growth to $[0,u_t]$, clipping ensures
$b_{t+1}=m\odot\operatorname{clip}(C_g\widetilde v_{t+1},0,u_t)$.
Thus $C_gv_{t+1}^H=c_{t+1}-c_t$, while the six exposures retain their
linearly propagated values. This correction makes the hybrid recurrence
nonlinear.

\subsection{Learned decoder: linear propagation of the observables}
\label{sec:decoder}
The learned decoder propagates
$v_{t+1}^D=L_Dv_t^D$ from $v_{t_0}^D=\mathcal V(X_{t_0})$, so
$v_{t_0+h}^D=L_D^h\mathcal V(X_{t_0})$.
Let $D\in\R^{7\times7}$ be the diagonal matrix of training root-mean-square
(RMS) observable values, defined in Section~\ref{sec:decoder-learning}.
With seven coefficients $\theta\in\R^7$ shared across cells, the decoder
computes the daily ignition hazard and probability as
\begin{equation}
 a_t(x)=\left[\theta^\top D^{-1}v_{t+1}^D(x)\right]_+,
 \label{eq:decoder-score}
\end{equation}
\begin{equation*}
 p_t^D=m\odot\left(1-\exp(-a_t)\right).
\end{equation*}
The positive part makes the hazard nonnegative. On burnable cells,
$\exp(-a_t)$ is the conditional probability of remaining unignited during
the day. We use $p_t^D$ in Equation~\eqref{eq:transition} and retain the
linearly propagated observables for the next step.

Let $\mathcal A_h=\sum_{j=0}^{h-1}a_{t_0+j}$ be the accumulated hazard.
The fixed binary mask and Equation~\eqref{eq:transition} give
\begin{equation}
 c_{t_0+h}=c_{t_0}+(1-c_{t_0})\odot
             \left[1-\exp(-m\odot\mathcal A_h)\right].
 \label{eq:hazard-sum}
\end{equation}
We first fit $L_D$ by ridge regression, then fit $\theta$ to multiday map
errors with $L_D$ and $D$ fixed (Section~\ref{sec:decoder-learning}).
Appendix~\ref{app:dynamics} analyzes the linear observable dynamics.

\section{Learning the operator and decoder}
\label{sec:learning}
This section fits the operator and the decoder.
Section~\ref{sec:regression} states the ridge regression for the
operator, Section~\ref{sec:selection} selects its two penalties within the
training segment, and Section~\ref{sec:decoder-learning} fits the seven
decoder coefficients.

\subsection{Structured EDMD regression}
\label{sec:regression}
Let $n$ be the number of training cell-transition pairs. EDMD regresses
the next-day observables of each pair on its filtered observables
$[T_kv_t](x)$, $k=0,\ldots,21$. We call these $154$ values the regressors
of the pair. Let $\Psi_B\in\R^{n\times49}$ contain one row per pair with
the regressors for $k=0,\ldots,6$, and let $\Psi_X\in\R^{n\times105}$
contain those for $k=7,\ldots,21$. Let $Y\in\R^{n\times7}$ contain the
observed next-day observables from Equation~\eqref{eq:lift}. For per-event
fitting, each row has weight $1/n$.
For pooled fitting over $F=6$ fires, a row from fire $\ell$ has weight
$1/(F n_\ell)$, where $n_\ell$ is its row count. Each fire therefore
contributes equally, regardless of its grid size. Let $\Pi$ be the diagonal matrix of these row weights, and let
$\langle A,B\rangle_\Pi=A^\top\Pi B$.

We scale the columns of $\Psi_B$ and $\Psi_X$ and orthogonalize the
latter against the former as detailed in Appendix~\ref{app:design}. Let $Z\in\R^{n\times154}$ be the resulting regressor matrix.
Let $\Lambda$ be diagonal, with $\lambda$ on its first 49 diagonal
entries and $\lambda_X$ on the remaining 105.
For the coefficient matrix $B\in\R^{154\times7}$, we solve
\begin{equation}
 \widehat B=\arg\min_B\left\{\operatorname{tr}[(ZB-Y)^\top\Pi(ZB-Y)]
                  +\operatorname{tr}(B^\top\Lambda B)\right\}
     =(Z^\top\Pi Z+\Lambda)^{-1}Z^\top\Pi Y.
 \label{eq:ridge}
\end{equation}
Undoing the transformations and reshaping $\widehat B$ gives the
matrices $W_k$ of the fitted operator $L_D$.

\subsection{Historical selection and final fits}
\label{sec:selection}

Residualization makes the normal matrix in Equation~\eqref{eq:ridge}
block diagonal. We select the box-filter penalty $\lambda$ first, then
select $\lambda_X$ with $\lambda$ fixed to assess the contribution of the
Gaussian-derivative filters. Both stages fit on $\mathcal P_8$ and score
forecasts from $d_8$, $d_9$, and $d_{10}$ against observed targets through
$d_{11}$ (Table~\ref{tab:selection}). We average pooled scores equally
over fires and break ties in favor of the larger penalty.

For a grid field $\zeta$, let
$\operatorname{area}(\zeta)=\sum_{x\in\Omega}\zeta(x)$. Let
$\widehat g_{t+1}$ be the predicted growth component and
$\widehat c_{t+1}=c_t+\widehat g_{t+1}$. The score of the one-day forecast
from $t$ is
\begin{equation}
 s_\lambda(t)=
 \frac{|\operatorname{area}(\widehat c_{t+1}-c_{t+1})|+
       |\operatorname{area}(\widehat g_{t+1}-g_{t+1})|}
      {2\max(10,\operatorname{area}(g_{t+1}))}.
 \label{eq:base-selection}
\end{equation}

For $\lambda_X$, we score the mean squared error (MSE) of the
uncorrected accumulated prediction
\begin{equation}
 \widetilde c_{t+h}=c_t+\sum_{j=1}^{h}C_g L^j\mathcal V(X_t)
 \label{eq:selection-accumulator}
\end{equation}
over the cells with $c_t=0$ at horizons $h\in\{1,3\}$. We average over
starts within each horizon and then over horizons. The candidate
$\lambda_X=\infty$ removes the Gaussian-derivative coefficients.

\begin{table}[H]\centering\small
\caption{Historical penalty selection within the training segment.}
\label{tab:selection}
\begin{tabular}{l>{\raggedright\arraybackslash}p{0.24\linewidth}>{\raggedright\arraybackslash}p{0.22\linewidth}>{\raggedright\arraybackslash}p{0.30\linewidth}}\toprule
Penalty & Candidates / filters & Eligible horizons & Selection score\\\midrule
$\lambda$ & $10^{-6},10^{-3},10^{-1}$ / 7 & One calendar day & Mean of Equation~\eqref{eq:base-selection}\\
$\lambda_X$ & $0.01,0.1,1,\infty$ / 22 & One and three days, target by $d_{11}$ & Raw map MSE from Equation~\eqref{eq:selection-accumulator}\\
\bottomrule\end{tabular}
\end{table}

After selection, we recompute the normalization and residualization from
the training data through $d_{11}$ and solve Equation~\eqref{eq:ridge} to
obtain $L_D$. 

\subsection{Learning seven decoder coefficients}
\label{sec:decoder-learning}
We normalize the observables by their training RMS values. For a per-event
fit, the diagonal entries of $D$ are
\begin{equation*}
 D_{jj}=\Big(\frac1{N|\mathcal P_{11}|}\sum_{t\in\mathcal P_{11}}
 \sum_{x\in\Omega}[v_{t-1}(x)]_j^2\Big)^{1/2},
\end{equation*}
where $v_{t-1}=\mathcal V(X_{t-1})$. For pooled fitting, $D_{jj}^2$
averages these mean squares equally over the six fires.

For each historical start with a preceding daily map, we propagate the
operator fitted through $d_{11}$ for one to five calendar days. Let
$\mathcal H_\ell$ contain the start-horizon pairs $(t,h)$ whose targets
are observed by $d_{11}$ for fire $\ell$, and let
$\Omega_{\ell,t}=\{x:c_{\ell,t}(x)=0\}$.
Let $\widehat c^\theta$ be the forecast of Equation~\eqref{eq:hazard-sum}.
We define the anchor $\theta_0=De_1$, where $e_1$ is the first standard
basis vector of $\R^7$. This anchor uses the positive latent growth as
the ignition hazard.
For $F=1$ in per-event fitting and $F=6$ in pooled fitting, we minimize
\begin{equation*}
 \mathcal J(\theta)=\frac1F\sum_{\ell=1}^F\frac1{|\mathcal H_\ell|}
 \sum_{(t,h)\in\mathcal H_\ell}\frac1{|\Omega_{\ell,t}|}
 \sum_{x\in\Omega_{\ell,t}}
 \left(\widehat c^\theta_{\ell,t+h}(x)-c_{\ell,t+h}(x)\right)^2
 +10^{-4}\|\theta-\theta_0\|_2^2.
\end{equation*}
We initialize at $\theta_0$ and run L-BFGS for at most 100 iterations,
keeping $L_D$, $D$, and the penalty fixed. All targets lie in the training
segment. The fitted coefficients remain fixed during continuous forecasts.

\section{Training and evaluation protocol}
\label{sec:protocol}
This section defines the three forecast protocols and, in
Section~\ref{sec:scores}, the probability scores and the IoU.

We compare per-event and pooled fits under the three protocols in
Table~\ref{tab:protocols}. Every predictor starts from the observed map
pair and cell state in Equation~\eqref{eq:observed-state}. Continuous
forecasts advance daily from $t_0=d_{11}$ to the end of the record.
We reserve $t_0+1$ and $t_0+2$ for threshold calibration and score only
later observed dates. Appendix~\ref{app:refitting} gives the historical
calibration and matched cases for daily restarts.

\begin{table}[H]\centering\small
\caption{Forecast protocols. All fits initially use maps through $d_{11}$.
In pooled expanding fits, only the current fire's history grows. The other
five retain their initial training segments and equal weights.}
\label{tab:protocols}
\begin{tabular}{>{\raggedright\arraybackslash}p{0.19\linewidth}>{\raggedright\arraybackslash}p{0.23\linewidth}>{\raggedright\arraybackslash}p{0.23\linewidth}>{\raggedright\arraybackslash}p{0.24\linewidth}}\toprule
Protocol & Initialization & Fitted coefficients & Scored targets\\\midrule
Continuous & Observed map pair at $d_{11}$ once & Fixed through $d_{11}$ & Observed dates from $d_{11}+3$\\
Daily restart, frozen & Latest observed map pair & Fixed through $d_{11}$ & One through five calendar days\\
Daily restart, expanding & Latest observed map pair & Refit through current start & One through five calendar days\\
\bottomrule\end{tabular}
\end{table}

\subsection{Probability scores and IoU}
\label{sec:scores}
For fire $\ell$ and forecast start $t_0$, we score only
$\Omega_{\ell,t_0}=\{x:c_{\ell,t_0}(x)=0\}$.
On a target date, let $y\in\{0,1\}$ indicate observed cumulative burning
and let $\pi=1-u_t$ be its predicted probability. This equals $b_t+r_t$
for our predictors and the sum of burning and burned probabilities for
CA--CNN. We report the Brier loss, also called MSE, and the log loss:
$L_B(\pi,y)=(\pi-y)^2$ and
$L_{\log}(\pi,y)=-y\log\pi-(1-y)\log(1-\pi)$~\cite{gneiting2007}.
We use average precision (AP) and the area under the receiver operating
characteristic curve (AUC) to measure spatial ranking.

For continuous forecasts, we classify a cell as burned when $\pi>\tau$.
We select $\tau$ to maximize mean IoU on the two calibration dates and
freeze it for testing. Each per-event model has its own threshold. Each
pooled model uses one threshold selected with equal fire weights.
Let $P_{\ell,t}$ and $Y_{\ell,t}$ be the predicted and observed burned
cells in $\Omega_{\ell,t_0}$, and let $\mathcal T_\ell$ be the test dates.
We compute
\begin{equation*}
 \mathcal I_{\ell,t}=\frac{|P_{\ell,t}\cap Y_{\ell,t}|}{|P_{\ell,t}\cup Y_{\ell,t}|},\qquad
 \overline{\mathcal I}_\ell=\frac1{|\mathcal T_\ell|}\sum_{t\in\mathcal T_\ell}\mathcal I_{\ell,t},\qquad
 \overline{\mathcal I}=\frac16\sum_{\ell=1}^6\overline{\mathcal I}_\ell.
\end{equation*}
All continuous-forecast metrics use this averaging over dates and then
fires. Persistence predicts $c_t=c_{t_0}$, giving zero probability of
additional burning on every scored cell.

\section{Forecast results}
\label{sec:results}
This section reports the selected penalties and four results: the
aggregate probability scores, the IoU by fire, the admissibility of the
forecasts, and the daily restarts.

We select $\lambda=10^{-3}$ and $\lambda_X=0.01$ for the pooled fit. The
per-event $\lambda$ is $10^{-6}$ for Bear, Brattain, and Pier and $0.1$ for
the other three fires. The per-event $\lambda_X$ is infinite for Bear and
Buck, $0.1$ for Chimney and Ferguson, and $0.01$ for Brattain and Pier. The
per-event operators of Bear and Buck therefore have 343 coefficients, and
the other operators 1,078.

\paragraph{Result 1: Both predictors improve aggregate probability scores.}
The learned decoder achieves the lowest aggregate MSE and log loss
(Table~\ref{tab:aggregate}). Both EDMD predictors improve MSE over
persistence on all six fires and over CA--CNN on four of six in each
fitting category. CA--CNN achieves the lower MSE on Chimney and Ferguson.
The decoder's aggregate advantage over the hybrid comes from Ferguson
and Pier, and per event also from Buck.

\begin{table}[H]\centering\small
\caption{Continuous-forecast scores, averaged over test dates within each fire and then equally over fires. We exclude the initial burned area and optimize IoU thresholds on separate calibration dates. Bold values indicate the best result for each metric within the same fitting category and evaluation cases. Ties at the displayed precision share bold type.}
\label{tab:aggregate}
\setlength{\tabcolsep}{4pt}
\begin{tabular}{llrrrrr}\toprule
Fit & Predictor & MSE $\downarrow$ & Log loss $\downarrow$ & AP $\uparrow$ & AUC $\uparrow$ & IoU $\uparrow$\\\midrule
Per-event & Hybrid EDMD & 0.0705 & 0.2893 & \textbf{0.6381} & 0.9125 & \textbf{0.4018}\\
Per-event & Learned decoder & \textbf{0.0685} & \textbf{0.2709} & 0.6279 & \textbf{0.9132} & 0.3863\\
Per-event & CA--CNN & 0.1601 & 1.6784 & 0.3876 & 0.8348 & 0.3717\\
Both & Persistence & 0.1123 & 1.8108 & 0.1123 & 0.5000 & 0.0000\\
Pooled & Hybrid EDMD & 0.0712 & 0.2900 & \textbf{0.6262} & 0.8999 & 0.3759\\
Pooled & Learned decoder & \textbf{0.0703} & \textbf{0.2812} & 0.6260 & \textbf{0.9024} & 0.3737\\
Pooled & CA--CNN & 0.1262 & 0.8886 & 0.4666 & 0.8423 & \textbf{0.3926}\\
\bottomrule\end{tabular}
\end{table}

\begin{figure}[H]\centering
\includegraphics[width=0.85\linewidth]{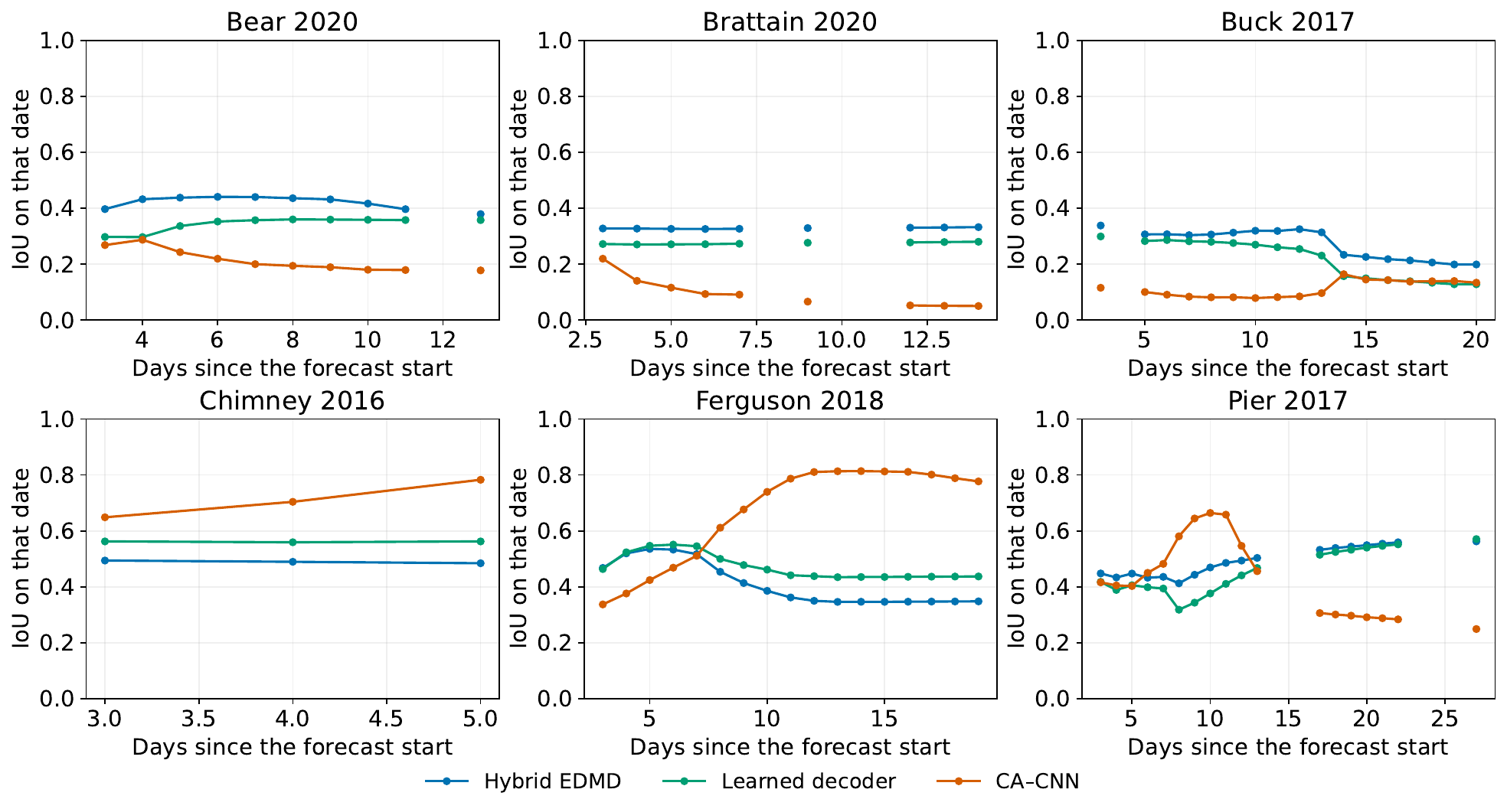}
\caption{Per-event IoU on each observed test date of the continuous
forecasts from $d_{11}$. Gaps indicate missing maps.}
\label{fig:iou-event}
\end{figure}

\begin{figure}[H]\centering
\includegraphics[width=0.85\linewidth]{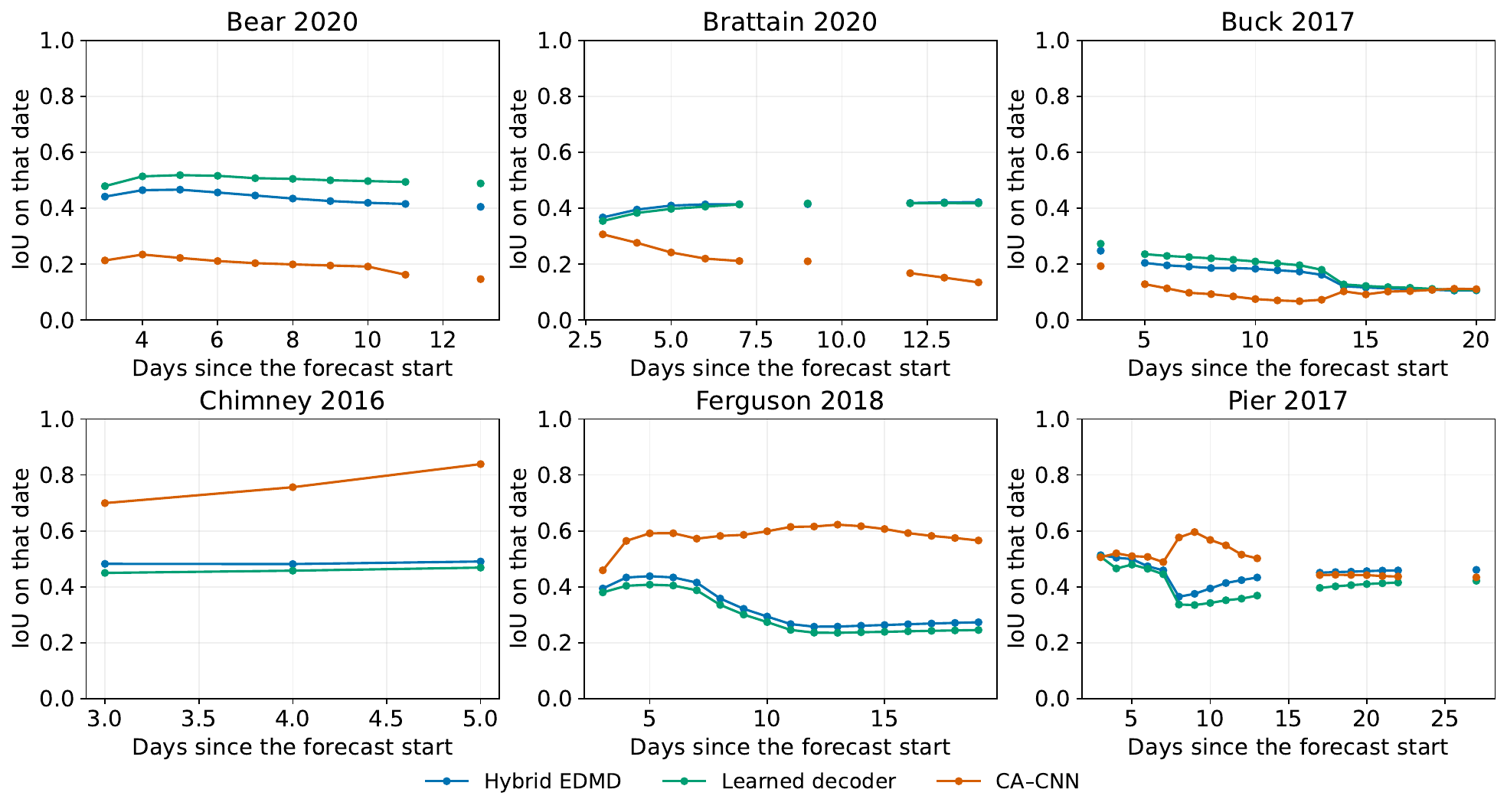}
\caption{Pooled IoU on the same test dates as Figure~\ref{fig:iou-event}.
Appendix~\ref{app:maps} shows the binary predictions on every observed date
from day $t_0+1$ to day $t_0+6$.}
\label{fig:iou-pooled}
\end{figure}

\paragraph{Result 2: The hybrid leads per-event IoU, and CA--CNN leads pooled IoU.}
Tables~\ref{tab:per_event} and~\ref{tab:pooled} show the variation across
fires. Each EDMD predictor exceeds CA--CNN's IoU on four fires per event
and three pooled. Figures~\ref{fig:iou-event} and~\ref{fig:iou-pooled}
show the daily comparisons. On Bear (Figure~\ref{fig:map-bear_2020}), both
EDMD predictors reduce the large false-positive regions of CA--CNN.
The per-event decoder also misses part of the right-hand front.

\begin{table}[H]\centering\small
\caption{Per-event continuous forecasts by fire. Entries average the indicated test dates using the protocol of Table~\ref{tab:aggregate}. Bold values indicate the best result for each metric within the same fitting category and evaluation cases. Ties at the displayed precision share bold type.}
\label{tab:per_event}
\setlength{\tabcolsep}{4pt}
\begin{tabular}{lrrrrrrrr}\toprule
Fire & Dates & \multicolumn{1}{c}{Persistence} & \multicolumn{2}{c}{Hybrid EDMD} & \multicolumn{2}{c}{Learned decoder} & \multicolumn{2}{c}{CA--CNN}\\
\cmidrule(lr){3-3}\cmidrule(lr){4-5}\cmidrule(lr){6-7}\cmidrule(lr){8-9}
 & & MSE $\downarrow$ & MSE $\downarrow$ & IoU $\uparrow$ & MSE $\downarrow$ & IoU $\uparrow$ & MSE $\downarrow$ & IoU $\uparrow$\\\midrule
Bear & 10 & 0.1169 & \textbf{0.0739} & \textbf{0.4203} & 0.0874 & 0.3428 & 0.2453 & 0.2132\\
Brattain & 9 & 0.0309 & \textbf{0.0209} & \textbf{0.3280} & 0.0232 & 0.2740 & 0.2685 & 0.0971\\
Buck & 17 & 0.0557 & 0.0453 & \textbf{0.2727} & \textbf{0.0447} & 0.2169 & 0.2040 & 0.1109\\
Chimney & 3 & 0.1017 & 0.0461 & 0.4892 & 0.0491 & 0.5613 & \textbf{0.0352} & \textbf{0.7119}\\
Ferguson & 17 & 0.2909 & 0.2004 & 0.4096 & 0.1707 & 0.4705 & \textbf{0.1071} & \textbf{0.6682}\\
Pier & 18 & 0.0780 & 0.0364 & \textbf{0.4913} & \textbf{0.0362} & 0.4522 & 0.1008 & 0.4288\\
\bottomrule\end{tabular}
\end{table}

\begin{table}[H]\centering\small
\caption{Pooled continuous forecasts by fire. Entries average the indicated test dates using the protocol of Table~\ref{tab:aggregate}. Bold values indicate the best result for each metric within the same fitting category and evaluation cases. Ties at the displayed precision share bold type.}
\label{tab:pooled}
\setlength{\tabcolsep}{4pt}
\begin{tabular}{lrrrrrrrr}\toprule
Fire & Dates & \multicolumn{1}{c}{Persistence} & \multicolumn{2}{c}{Hybrid EDMD} & \multicolumn{2}{c}{Learned decoder} & \multicolumn{2}{c}{CA--CNN}\\
\cmidrule(lr){3-3}\cmidrule(lr){4-5}\cmidrule(lr){6-7}\cmidrule(lr){8-9}
 & & MSE $\downarrow$ & MSE $\downarrow$ & IoU $\uparrow$ & MSE $\downarrow$ & IoU $\uparrow$ & MSE $\downarrow$ & IoU $\uparrow$\\\midrule
Bear & 10 & 0.1169 & \textbf{0.0728} & 0.4370 & 0.0808 & \textbf{0.5015} & 0.1936 & 0.1973\\
Brattain & 9 & 0.0309 & \textbf{0.0171} & \textbf{0.4079} & 0.0180 & 0.4023 & 0.0770 & 0.2128\\
Buck & 17 & 0.0557 & \textbf{0.0495} & 0.1577 & 0.0504 & \textbf{0.1757} & 0.2720 & 0.1008\\
Chimney & 3 & 0.1017 & 0.0499 & 0.4846 & 0.0550 & 0.4586 & \textbf{0.0225} & \textbf{0.7649}\\
Ferguson & 17 & 0.2909 & 0.1963 & 0.3217 & 0.1777 & 0.2975 & \textbf{0.1415} & \textbf{0.5844}\\
Pier & 18 & 0.0780 & 0.0418 & 0.4468 & \textbf{0.0398} & 0.4064 & 0.0503 & \textbf{0.4952}\\
\bottomrule\end{tabular}
\end{table}

\paragraph{Result 3: Both predictors keep the cell-state probabilities admissible.}
The hybrid forecasts satisfy the bounds and the monotonicity at every step.
For the learned decoder, the largest violation of the simplex or transition
constraints over all daily steps is $1.11\times10^{-15}$, the float64
round-off. The fuel map classifies as nonburnable some cells that later
burn, from 2 on Chimney to 3,788 on Ferguson. Both predictors assign these
cells zero probability, and the log loss penalizes them heavily. We keep
them in the scores, so the scores include the cost of these fuel-map
errors.


\paragraph{Result 4: With daily restarts, the hybrid attains the lower MSE at one and three days.}
Appendix~\ref{app:refitting} repeats the comparison with forecasts that
restart from each new map. With frozen coefficients, the hybrid attains the
lower MSE at one and three days. At five days, the per-event decoder and
the pooled hybrid attain the lower MSE. Both predictors attain a lower MSE
than CA--CNN at every horizon. Refitting on all available
maps reduces the MSE of the per-event decoder by 4.8\% at one day and
1.9\% at five days, so new maps improve its short-term forecasts most.

\section{Discussion}
\label{sec:discussion}
We first relate our method to Koopman and fire-spread models. We then
discuss the linear dynamics of the learned decoder and the computational
cost, and close with the limitations of the study.

Our method combines spatial Koopman approximation with admissible
cell-state transitions. Koopman analysis of PDEs treats observables as
functionals of spatial fields~\cite{nakao2020}, which motivates our use of
neighborhood observables. For translation-equivariant dynamics, local
Koopman approximations can share coefficients across a
domain~\cite{peitz2025}. We adopt spatial sharing as a modeling assumption
to keep the coefficient count independent of grid size. Physics-informed
DMD restricts the fitted matrix to encode properties such as locality and
shift equivariance~\cite{baddoo2023}. Our prescribed filter basis imposes
these spatial restrictions, while Equation~\eqref{eq:transition} enforces
the cell-state constraints. Reprojection methods map EDMD predictions back onto the manifold of
observable values realizable by physical states~\cite{vangoor2023}.
Our hybrid applies a partial consistency correction: it replaces predicted
growth with accepted growth and retains the propagated exposures
(Section~\ref{sec:hybrid}). Full reprojection would also reconstruct the
exposures from the updated maps. Neural Koopman models learn encoders and nonlinear decoders from
trajectories~\cite{lusch2018}, and Koopman neural operators learn PDE
solution operators through latent Koopman
approximations~\cite{xiong2024}. We fix seven physically defined
observables, fit their propagation by ridge regression, and learn seven
hazard coefficients for the decoder. A related study learns
Koopman-inspired Fourier and convolutional operators for simulated
Michelson--Sivashinsky and Kuramoto--Sivashinsky flame-front
dynamics~\cite{flame2025}. Its two-dimensional Fourier model uses 45
training sequences per case and about 43 hours on an A100. Our daily
wildfire forecasts use eleven observed maps per fire. The different
prediction tasks and training data preclude a direct accuracy or runtime
ranking between these studies.

CA--CNN learns spatial spread parameters through a CNN~\cite{cacnn},
while PyTorchFire calibrates a differentiable cellular automaton by
gradient descent~\cite{pytorchfire}. Both evolve fire states through local
ignition rules. Our decoder instead computes ignition from propagated
observables, using CA--CNN's exponential hazard-to-probability conversion
at a daily interval. A workshop report applies DMD to annual satellite
brightness maps for regional fire-occurrence prediction~\cite{firerisk}.
A workshop poster describes ongoing EDMD modeling of smoke plumes from
WRF--SFIRE simulations~\cite{smokeedmd}. These applications concern
regional occurrence and smoke transport, while we forecast the daily
burned-area progression of individual fires.

The learned decoder also permits direct analysis of its linear observable
dynamics (Appendix~\ref{app:dynamics}). The finite-grid latent norms fall
to at most 15.6\% of their initial values after six days. Pooling slows
this decay on four fires. The Fourier analysis gives spectral radii below
one for the corresponding unbounded-grid operators. These quantities
describe the latent dynamics. The decoder and remaining unburned area
jointly determine the physical growth, which also declines in the saved
forecasts.

The pooled hybrid has 1,078 coefficients, and the decoder adds seven,
against 36,018 CA--CNN parameters. On an RTX 4090, accumulating regression
statistics for six fires took 16.0\,s, solving all frozen and expanding
fits took 4.9\,s, and fitting seven decoders took 5.5\,s. On a 32-thread
Intel Core i9-14900KF CPU, the same accumulation took 56\,s and a six-day
forecast on the largest grid took 10\,s. For joint six-fire training, the
CA--CNN repository reports 10 to 15\,h on one H200 for 5,000
epochs~\cite{cacnncode}. The small EDMD training cost supports frequent refitting.

Several limitations define the scope of these results. We used these six
fires to choose the observables and operator structure, and pooled
training includes an initial segment of every evaluated fire. Transfer
requires evaluation on additional fires with a fixed design. The CA--CNN
comparison uses supplied checkpoints whose training and calibration
procedures differ from ours.

Environmental inputs did not improve historical validation. We conditioned
the filter responses on terrain, fuel, and weather. We used group
lasso~\cite{yuan2006} to select entire coefficient blocks, then refitted the
selected blocks by ridge regression.
Validation selected the operator without environmental blocks. A separate
controlled-wind experiment on Bear found weak typical sensitivity.
Recent growth may already encode environmental effects, and the short
records may limit estimation of additional coefficients.

The fuel mask prevents ignition within a lake, but the spatial filters
can transmit influence across it. Preventing such paths would require
barrier-aware spatial coupling. Both predictors also remain at zero
growth when initialized with zero recent growth. Predicting renewed spread
would require additional information or a source term. Finally, the
conditional-mean interpretation relies on the Markov and closure
assumptions of Section~\ref{sec:koopman}.

\section{Conclusion}
\label{sec:conclusion}
Spatial EDMD provides compact daily wildfire predictors with admissible
cell states. Hybrid feedback and a seven-coefficient learned decoder both
improve aggregate probability scores over the CA--CNN. The hybrid achieves the highest per-event IoU, while pooled
CA--CNN leads pooled IoU. Daily restarts preserve the MSE advantage over
CA--CNN, and expanding-history refits modestly improve the per-event
decoder. Training in seconds makes these predictors practical to update
as daily maps arrive. Their performance across six fires supports further
evaluation on independent events, especially for renewed growth and
propagation across fuel barriers.

\appendix
\section{Propagation kernels and regressor matrix}
\label{app:mechanics}
Appendix~\ref{app:kernels} defines the Gaussian-derivative filters of
Section~\ref{sec:filters}, and Appendix~\ref{app:design} constructs the
regressor matrix $Z$ of Equation~\eqref{eq:ridge}.

\subsection{Signed spatial kernels}
\label{app:kernels}
For a scale $\sigma$, let $\delta_\sigma=\min(31,\lceil3\sigma\rceil)$.
For integer offsets $j$ with $|j|\le\delta_\sigma$, we define
\begin{align*}
 \eta^{(0)}_\sigma(j)&=\frac{\exp[-j^2/(2\sigma^2)]}{\sum_{|i|\le\delta_\sigma}\exp[-i^2/(2\sigma^2)]},\\
 \eta^{(1)}_\sigma(j)&\propto j\eta^{(0)}_\sigma(j),\qquad
 \eta^{(2)}_\sigma(j)\propto
 \left(j^2-\sum_{|i|\le\delta_\sigma}i^2\eta^{(0)}_\sigma(i)\right)\eta^{(0)}_\sigma(j).
\end{align*}
and scale $\eta^{(1)}_\sigma$ and $\eta^{(2)}_\sigma$ to unit absolute
sum. The next display omits the scale subscript. At each scale, the five
separable kernels are
\begin{equation*}
 \eta^{(0)}\otimes\eta^{(0)},\quad
 \eta^{(0)}\otimes\eta^{(1)},\quad
 \eta^{(1)}\otimes\eta^{(0)},\quad
 \tfrac12(\eta^{(0)}\otimes\eta^{(2)}-\eta^{(2)}\otimes\eta^{(0)}),\quad
 \eta^{(1)}\otimes\eta^{(1)}.
\end{equation*}
The first factor acts along the north-south axis and the second along the
west-east axis. The scales $\sigma=2$, $8$, and $16$ give the filters
$T_7$ to $T_{21}$ in this order. A filter with kernel $\kappa$ acts on a
scalar field $\zeta$ by correlation,
\begin{equation}
 [T\zeta](x)=\sum_\delta\kappa(\delta)\zeta(x+\delta),
 \label{eq:correlation}
\end{equation}
with $\zeta=0$ outside $\Omega$. Every filter reaches at most 31 cells in each
direction, so zero padding changes a filtered value only within 31 cells
of the grid edge.

\subsection{Construction of the regressor matrix}
\label{app:design}
For a matrix $A$ with $n$ rows, let $\operatorname{rms}_\Pi(A)$ be the
diagonal matrix of the square roots of the diagonal entries of
$\langle A,A\rangle_\Pi$. Its entries are the weighted RMS values of the
columns of $A$. We construct $Z$ in two steps. First, we scale every column to unit weighted RMS, so that $\lambda$ and
$\lambda_X$ penalize coefficients of columns with equal magnitude. Recall that the
rows of $\Psi_B$ and $\Psi_X$ contain the filtered observables
$[T_kv_t](x)$ for $k=0,\ldots,6$ and $k=7,\ldots,21$, respectively
(Section~\ref{sec:learning}). With
$S_B=\operatorname{rms}_\Pi(\Psi_B)$ and
$S_X=\operatorname{rms}_\Pi(\Psi_X)$, let
$\overline\Psi_B=\Psi_BS_B^{-1}$ and $\overline\Psi_X=\Psi_XS_X^{-1}$. Second, we regress each column of $\overline\Psi_X$ on the columns of
$\overline\Psi_B$ by weighted least squares and keep the rescaled residual:
\begin{equation*}
 \Gamma=\langle\overline\Psi_B,\overline\Psi_B\rangle_\Pi^\dagger
              \langle\overline\Psi_B,\overline\Psi_X\rangle_\Pi,\qquad
 R_X=(\overline\Psi_X-\overline\Psi_B\Gamma)S_\perp^{-1}.
\end{equation*}
The matrix $\Gamma\in\R^{49\times105}$ contains the regression
coefficients, and
$S_\perp=\operatorname{rms}_\Pi(\overline\Psi_X-\overline\Psi_B\Gamma)$.
Hence $\langle\overline\Psi_B,R_X\rangle_\Pi=0$. We set
$Z=[\overline\Psi_B,R_X]$. The matrix $Z^\top\Pi Z+\Lambda$ of
Equation~\eqref{eq:ridge} is then block diagonal, and this property gives
the two-stage fit of Section~\ref{sec:selection}.


\section{Linear dynamics of the learned decoder}
\label{app:dynamics}
The observables of the learned decoder evolve linearly,
$v_{t_0+j}^D=L_D^jv_{t_0}^D$ (Section~\ref{sec:decoder}). This appendix
shows how the powers of $L_D$ determine the forecast burned area and
measures their growth and decay. The hybrid replaces the predicted growth after every
step, so powers of $L_D$ do not describe its forecast.

\paragraph{Response of the burned area.}
Fix $c_{t_0}$ and $m$. For a quantity $f$ that depends on the initial
observables $v_{t_0}^D$, let $\delta f$ be its derivative with respect to
$v_{t_0}^D$ in the direction $\delta v$. By
Equation~\eqref{eq:hazard-sum},
$u_{t_0+h}=1-c_{t_0+h}=(1-c_{t_0})\odot\exp(-m\odot\mathcal A_h)$. The
chain rule applied to Equation~\eqref{eq:hazard-sum} therefore gives
\begin{equation*}
 \delta c_{t_0+h}=(1-c_{t_0})\odot\exp(-m\odot\mathcal A_h)\odot m\odot
 \delta\mathcal A_h
 =u_{t_0+h}\odot m\odot\sum_{j=1}^h\delta a_{t_0+j-1}.
\end{equation*}
By Equation~\eqref{eq:decoder-score}, $a_{t_0+j-1}(x)$ is the positive
part of the linear function $\theta^\top D^{-1}[L_D^jv_{t_0}^D](x)$. Where
this function is nonzero,
$\delta a_{t_0+j-1}(x)=\chi_j(x)\,\theta^\top D^{-1}[L_D^j\delta v](x)$,
with $\chi_j(x)=1$ if $a_{t_0+j-1}(x)>0$ and $\chi_j(x)=0$ otherwise.
Hence
\begin{equation}
 \delta c_{t_0+h}(x)=u_{t_0+h}(x)m(x)
 \sum_{j=1}^h\chi_j(x)\,\theta^\top D^{-1}[L_D^j\delta v](x).
 \label{eq:decoded-sensitivity}
\end{equation}
At each cell, the response sums the terms $\theta^\top D^{-1}[L_D^j\delta v](x)$
over the days with positive hazard and weights them by the unburned,
burnable fraction $u_{t_0+h}m$.
The growth or decay of $L_D^j$ with $j$ therefore determines how strongly
the initial observables affect the forecast at horizon $h$.

\paragraph{Scaled operator.}
The seven observables have different magnitudes. We therefore scale them
by one fixed matrix $D_*$, the matrix $D$ of
Equation~\eqref{eq:decoder-score} computed for the pooled fit. We use $D_*$
for every operator, including the per-event operators, so that all gains
below are comparable. Let $\overline Lv=D_*^{-1}L_D(D_*v)$ and
$\overline v_{t_0}=D_*^{-1}v_{t_0}^D$, where $D_*$ multiplies the
observables at each cell. The operators $\overline L$ and $L_D$ have the
same eigenvalues. 
all its values.

\paragraph{Frequency domain.}
Let $\kappa_k$ be the kernel of $T_k$ in Equation~\eqref{eq:correlation},
and apply the filters on the unbounded grid $\mathbb Z^2$. For a spatial frequency $\omega\in[-\pi,\pi]^2$ and
$a\in\mathbb C^7$, the operator $\overline L$ therefore maps the field
$e^{i\omega\cdot x}a$ to $e^{i\omega\cdot x}M(\omega)a$, with
\begin{equation*}
 M(\omega)=\sum_{k=0}^{21}\hat\kappa_k(\omega)\,D_*^{-1}W_kD_*,\qquad
 \hat\kappa_k(\omega)=\sum_\delta\kappa_k(\delta)e^{i\omega\cdot\delta}.
\end{equation*}
The Fourier transform thus turns $\overline L$ into multiplication by the
$7\times7$ matrix $M(\omega)$. Hence the spectrum of $\overline L$ on
$\mathbb Z^2$ is the union over $\omega$ of the eigenvalues of $M(\omega)$, and describes the dynamics in the interior of the
grid.

\paragraph{Spectral radius and realized gain.}
For a square matrix $A$, let $r(A)$ be the largest modulus of its
eigenvalues. We report two quantities. First, if $M(\omega)a=\mu a$, then
$\overline L^j$ maps $e^{i\omega\cdot x}a$ to $\mu^je^{i\omega\cdot x}a$.
By Equation~\eqref{eq:decoded-sensitivity}, such a perturbation enters the
response on day $t_0+j$ with the factor $\mu^j$. The spectral radius
$r_*=\max_\omega r(M(\omega))$ therefore sets the slowest asymptotic decay
on the unbounded grid.
We compute $r_*$ by maximizing over $\omega$ by a grid search followed by local optimization.
Second, the realized gain $\rho_h=\frac{\|\overline L^h\overline v_{t_0}\|_2}{\|\overline v_{t_0}\|_2}$, computed on the grid of the fire, is the amplification of the observed
initial observables.

\paragraph{Results.}
Table~\ref{tab:latent} lists these quantities. The pooled operators share
the matrices $W_k$ and hence $M(\omega)$, so they share $r_*$. Every operator satisfies $r_*<1$,
with values from 0.571 to 0.941, so the latent observables decay
asymptotically on the unbounded grid. On the grids of the fires, the
realized gains measure the decay of the actual forecasts, with
$\rho_6\le0.156$ for every operator. Pooling raises $\rho_6$ on four
fires and lowers it on Bear and Pier. The largest increases occur on
Brattain and Chimney. On Brattain, $\rho_6$ rises from 0.004 for the
per-event operator to 0.046 for the pooled operator. On Chimney, it rises
from 0.008 to 0.046. Pooled forecasts therefore keep predicting growth
for longer on these fires.

\paragraph{Fire-wide totals.}
On the unbounded grid, $\sum_x[T_k\zeta](x)=\hat\kappa_k(0)\sum_x\zeta(x)$,
so $\sum_x[\overline Lv](x)=M(0)\sum_xv(x)$. The fire-wide totals of the
latent observables then evolve by $M(0)$. On the grid of a fire, zero
padding removes part of the filtered mass near the edge. A normalized
$3\times3$ average, for example, keeps only $4/9$ of an impulse in a
corner cell. For the pooled operator and the forecasts from $d_{11}$, the
one-step relative error of this recursion ranges from 0.7\% to 8.7\% over
the first 20 days, so the recursion holds only approximately. The dominant
eigenvalue of $M(0)$ for the pooled operator is 0.868. It describes the
asymptotic decay of the idealized totals recursion when its dominant mode
is excited. Finite-grid forecasts also include boundary contributions. On
day 20, the ratio of consecutive latent growth totals ranges from 0.834 to
0.861 across the six fires.

These totals are latent quantities. For observables computed from maps,
the growth total equals the burned area added during the last day. After
propagation and decoding, the two differ. On the first forecast day of
pooled Bear, the latent growth total is 3,629 cells, whereas the decoder
adds 5,478 cells of burned area. The eigenvalue therefore describes the
latent totals, not the burned area. The predicted burned area nevertheless
declines at a similar rate. The forecasts of Buck, Ferguson, and Pier
cover 19 days or more. Their ratios of predicted new burned area on
consecutive days approach 0.855, 0.862, and 0.863.

\begin{table}[H]\centering\small
\caption{Spectral radius $r_*$ and realized gains $\rho_h$ of the scaled operator $\overline L$ of the learned decoder. The pooled operators share $r_*$.}
\label{tab:latent}
\setlength{\tabcolsep}{4pt}
\begin{tabular}{llrrrr}\toprule
Fire & Fit & $r_*$ & $\rho_1$ & $\rho_3$ & $\rho_6$\\\midrule
Bear & per-event & 0.873 & 0.462 & 0.211 & 0.078\\
Bear & pooled & 0.868 & 0.386 & 0.131 & 0.055\\
Brattain & per-event & 0.641 & 0.368 & 0.048 & 0.004\\
Brattain & pooled & 0.868 & 0.400 & 0.127 & 0.046\\
Buck & per-event & 0.775 & 0.344 & 0.113 & 0.036\\
Buck & pooled & 0.868 & 0.410 & 0.176 & 0.083\\
Chimney & per-event & 0.571 & 0.351 & 0.070 & 0.008\\
Chimney & pooled & 0.868 & 0.391 & 0.125 & 0.046\\
Ferguson & per-event & 0.774 & 0.418 & 0.137 & 0.046\\
Ferguson & pooled & 0.868 & 0.421 & 0.155 & 0.068\\
Pier & per-event & 0.941 & 0.446 & 0.234 & 0.156\\
Pier & pooled & 0.868 & 0.435 & 0.218 & 0.114\\
\bottomrule\end{tabular}
\end{table}

\section{Daily restarts and expanding-history refitting}
\label{app:refitting}
This appendix defines the daily-restart experiment, its two coefficient
policies, and its threshold calibration. It then reports the results for
frozen coefficients, the comparison with CA--CNN, expanding history, and
threshold updates.

In a second experiment, every forecast starts from the latest observed map.
We restart the hybrid, the learned decoder, and CA--CNN at every observed
date $d_i$ with $i\ge11$ and $d_i-1\in\mathcal D$. Each forecast covers
one to five days. A start also needs an observed target within five days
and a training target in $\mathcal P_i\cap(d_i-5,d_i]$. The second
condition excludes the Brattain start on 2020-09-30 and the Pier start on
2017-09-26. The six fires provide 80 starts and 320 observed pairs of start
and horizon (Table~\ref{tab:operational-counts}). These scores are
comparable only within this protocol.

We compare two coefficient policies. The frozen policy keeps the operator
fitted through $d_{11}$ and, for the learned decoder, its coefficients and
normalization. The expanding policy refits all coefficients and
normalizations on the maps available at each start. In pooled refits, only
the history of the current fire grows, and the other five fires keep their
training segments through $d_{11}$ and their equal weights. The filter
selection, the penalties, and the decoder objective stay fixed. Only the
coefficients and normalizations therefore change.

Thresholds use only observations available at the forecast start. We fit an
auxiliary model without the two latest one-day transitions of the current
fire, forecast these two transitions, and choose the threshold that
maximizes their mean IoU. With complete daily maps, we fit the auxiliary
model for a start $d_i$ through $d_{i-2}$ and forecast $d_{i-1}$ and $d_i$
from $d_{i-2}$ and $d_{i-1}$. With missing maps, we use the two latest
targets in $\mathcal P_i$ and fit through the preceding eligible target.
For the learned decoder, the auxiliary fits of the operator and the decoder
both end at the last remaining training target of the current fire. The
frozen policy selects its threshold once at $d_{11}$, and the expanding
policy at every start. Each fire and fitting category has one threshold for
all five horizons. CA--CNN keeps its trained checkpoints and receives the
same observed map pair at every start. It advances 50 microsteps per day
with date-aligned wind, and its threshold comes from the two transitions
that calibrate EDMD at $d_{11}$. Table~\ref{tab:calibration} summarizes
the timelines.

\begin{table}[H]\centering\small
\caption{Threshold calibration and forecast availability. Arrows indicate
chronological order. Each threshold is shared across the forecast horizons.}
\label{tab:calibration}
\begin{tabular}{>{\raggedright\arraybackslash}p{0.22\linewidth}>{\raggedright\arraybackslash}p{0.71\linewidth}}\toprule
Protocol & Fitting, calibration, and deployment timeline\\\midrule
Continuous & Fit through $d_{11}$ $\to$ initialize once at $d_{11}$ $\to$
calibrate on $d_{11}+1,d_{11}+2$ $\to$ score observed dates from $d_{11}+3$.
Probabilities originate at $d_{11}$. Binary decisions are available after
$d_{11}+2$. One threshold per event, or one shared across six fires for a pooled model.\\
Frozen restart & Withhold the last two daily targets when fitting the auxiliary EDMD model through
$d_{11}$ $\to$ calibrate on independent one-day forecasts of those targets
$\to$ deploy the full fit through $d_{11}$ and the fixed threshold at every eligible start.
One threshold per fire and fitting category. CA--CNN uses its trained checkpoint for calibration and deployment.\\
Expanding restart & At each $d_i$, withhold the last two daily targets when fitting the auxiliary model $\to$ calibrate on independent one-day forecasts $\to$ deploy a fit
through $d_i$. The threshold and binary forecasts are available at $d_i$.
One threshold per fire and fitting category.\\
\bottomrule\end{tabular}
\end{table}

\begin{table}[H]\centering\small
\caption{Daily observed restarts on matched forecast starts. Under the frozen policy, we retain the operator and decoder coefficients fitted through $d_{11}$. Under the expanding policy, we fit using all observations through each start. We exclude the current observed burned area and average scores over starts within each fire and horizon, then equally over fires. We retain the CA--CNN checkpoint and historical threshold at every restart. Bold values indicate the best result for each metric within the same fitting category and evaluation cases. Ties at the displayed precision share bold type.}
\label{tab:operational}
\setlength{\tabcolsep}{4pt}
\begin{tabular}{lllrrrrrr}\toprule
Fit & Predictor & History & 1-day MSE & IoU & 3-day MSE & IoU & 5-day MSE & IoU\\\midrule
Per-event & Hybrid & Frozen & 0.0089 & 0.2360 & \textbf{0.0220} & 0.2730 & 0.0358 & 0.2719\\
Per-event & Hybrid & Expanding & \textbf{0.0088} & \textbf{0.2479} & \textbf{0.0220} & 0.2871 & 0.0356 & 0.2820\\
Per-event & Decoder & Frozen & 0.0113 & 0.2316 & 0.0240 & 0.2883 & 0.0356 & 0.2833\\
Per-event & Decoder & Expanding & 0.0107 & 0.2361 & 0.0233 & 0.2957 & \textbf{0.0349} & 0.3113\\
Per-event & CA--CNN & Frozen & 0.0173 & 0.2159 & 0.0442 & \textbf{0.3289} & 0.0704 & \textbf{0.3583}\\
Pooled & Hybrid & Frozen & \textbf{0.0087} & 0.2458 & \textbf{0.0214} & 0.2756 & 0.0349 & 0.2667\\
Pooled & Hybrid & Expanding & \textbf{0.0087} & 0.2546 & \textbf{0.0214} & 0.3046 & \textbf{0.0348} & 0.3010\\
Pooled & Decoder & Frozen & 0.0094 & \textbf{0.2570} & 0.0219 & 0.2996 & 0.0355 & 0.2798\\
Pooled & Decoder & Expanding & 0.0093 & 0.2364 & 0.0218 & 0.3006 & 0.0353 & 0.2866\\
Pooled & CA--CNN & Frozen & 0.0134 & 0.2309 & 0.0309 & \textbf{0.3349} & 0.0476 & \textbf{0.3557}\\
\bottomrule\end{tabular}
\end{table}

\begin{figure}[H]\centering
\includegraphics[width=0.8\linewidth]{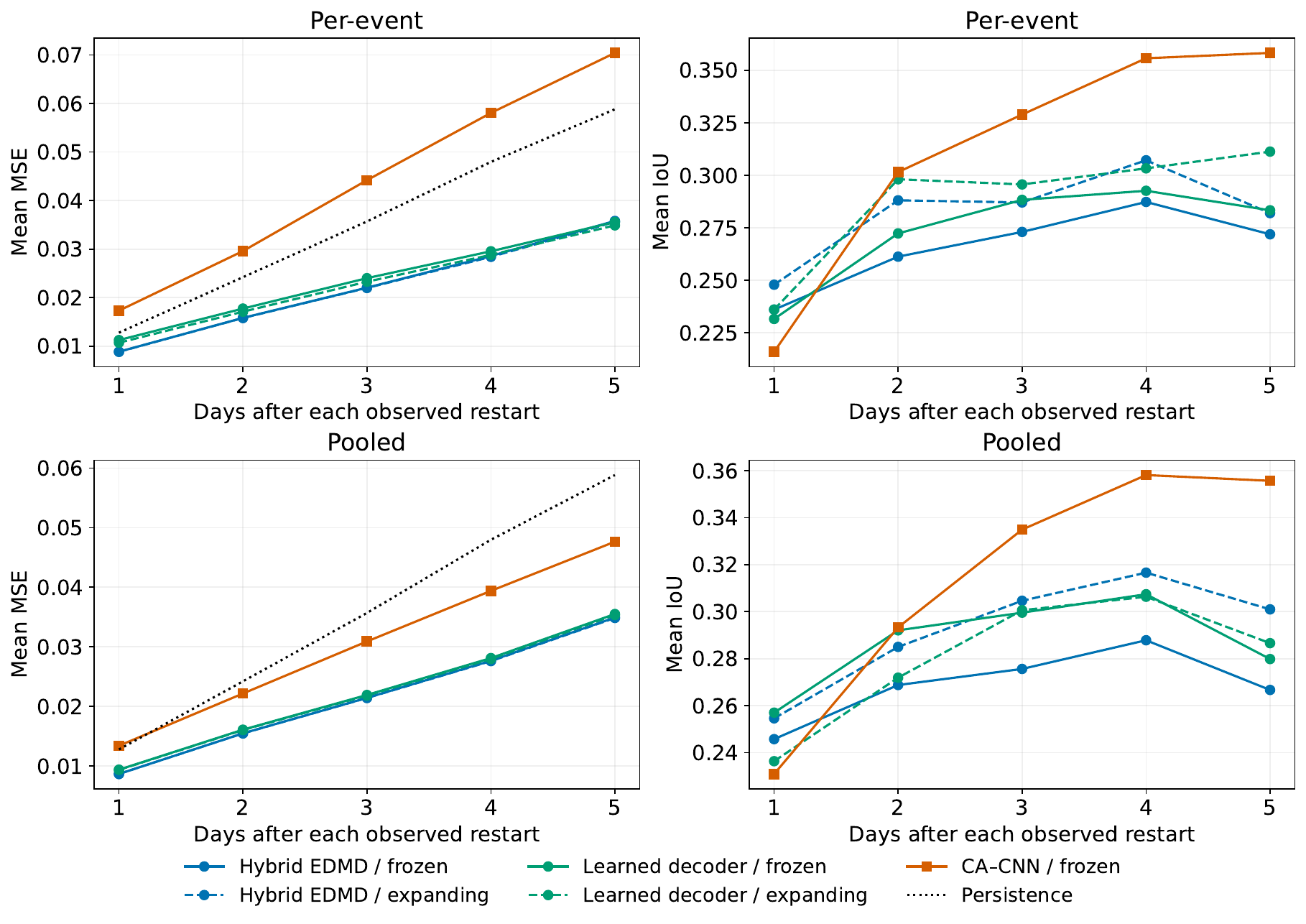}
\caption{Daily-restart MSE and IoU at all five horizons. Solid curves keep
the coefficients and threshold of $d_{11}$, and dashed curves refit both at
each start. Orange squares show CA--CNN. Persistence starts from the same
observed map.}
\label{fig:operational}
\end{figure}

\paragraph{Frozen coefficients.}
With frozen coefficients, the hybrid attains the lower MSE at one and three
days (Table~\ref{tab:operational} and Figure~\ref{fig:operational}). At
one day, the hybrid attains 0.0089 per event and 0.0087 pooled, against
0.0113 and 0.0094 for the decoder. At five days, the per-event decoder
attains 0.0356 against 0.0358 for the hybrid. The pooled hybrid attains
0.0349 against 0.0355 for the decoder. The pooled decoder attains the higher IoU at one, three, and
five days.

\paragraph{Comparison with CA--CNN.}
Both EDMD predictors attain a lower MSE than CA--CNN at every horizon. For
example, the five-day MSE of the frozen hybrid is 0.0358 per event and
0.0349 pooled. CA--CNN attains 0.0704 and 0.0476. CA--CNN attains the
higher IoU at three and five days, and the frozen hybrid at one day. Daily
restarts thus keep the advantage of EDMD in probability error, while
CA--CNN attains the larger thresholded overlap at longer horizons.

\paragraph{Expanding history.}
Refitting on all available maps reduces the MSE of the per-event decoder by
4.81\% at one day and 1.94\% at five days. The pooled reductions are only 0.37\%
to 0.47\%, so refitting matters mainly for the per-event decoder. With refitting, the ranking at each horizon matches that of the frozen
coefficients.

\paragraph{Threshold updates.}
To separate the two updates, Table~\ref{tab:crossed} also applies the
frozen threshold to the refitted forecasts and the updated threshold to the
frozen forecasts. For the pooled decoder at one day, updating both the
coefficients and the threshold lowers the IoU from 0.2570 to 0.2364, and
the threshold update accounts for most of this loss. At five days, updating
both raises the IoU from 0.2833 to 0.3113 per event and from 0.2798 to
0.2866 pooled. Appendix~\ref{app:operational} reports the results by fire.

\begin{table}[H]\centering\small
\caption{IoU with operator and decoder refitting separated from threshold updates. We restart every forecast from the current observation and update thresholds using past observations only. We average scores over starts within each fire and horizon, then equally over fires.}
\label{tab:crossed}
\setlength{\tabcolsep}{4pt}
\begin{tabular}{llllrrr}\toprule
Fit & Predictor & History & Threshold & 1 day & 3 days & 5 days\\\midrule
Per-event & Hybrid & Frozen & Frozen & 0.2360 & 0.2730 & 0.2719\\
Per-event & Hybrid & Frozen & Updated & 0.2363 & 0.2780 & 0.2815\\
Per-event & Hybrid & Expanding & Frozen & 0.2362 & 0.2751 & 0.2729\\
Per-event & Hybrid & Expanding & Updated & 0.2479 & 0.2871 & 0.2820\\
Per-event & Decoder & Frozen & Frozen & 0.2316 & 0.2883 & 0.2833\\
Per-event & Decoder & Frozen & Updated & 0.2320 & 0.2832 & 0.2944\\
Per-event & Decoder & Expanding & Frozen & 0.2137 & 0.2884 & 0.2810\\
Per-event & Decoder & Expanding & Updated & 0.2361 & 0.2957 & 0.3113\\
Pooled & Hybrid & Frozen & Frozen & 0.2458 & 0.2756 & 0.2667\\
Pooled & Hybrid & Frozen & Updated & 0.2548 & 0.3021 & 0.3013\\
Pooled & Hybrid & Expanding & Frozen & 0.2474 & 0.2771 & 0.2666\\
Pooled & Hybrid & Expanding & Updated & 0.2546 & 0.3046 & 0.3010\\
Pooled & Decoder & Frozen & Frozen & 0.2570 & 0.2996 & 0.2798\\
Pooled & Decoder & Frozen & Updated & 0.2374 & 0.2987 & 0.2865\\
Pooled & Decoder & Expanding & Frozen & 0.2567 & 0.3010 & 0.2811\\
Pooled & Decoder & Expanding & Updated & 0.2364 & 0.3006 & 0.2866\\
\bottomrule\end{tabular}
\end{table}

\section{Daily-restart results by fire}
\label{app:operational}
Table~\ref{tab:operational-counts} lists the matched daily-restart coverage.
Tables~\ref{tab:operational-per_event} and~\ref{tab:operational-pooled}
report the per-event and pooled results by fire. All predictors and both
policies share the same starts and targets, so their scores are directly
comparable. Chimney has one five-day start,
at $d_{11}$, so both policies give the same five-day score there.
\begin{table}[H]\centering\small
\caption{Matched daily-restart coverage. Each horizon column counts observed targets. For each start, we require a previous-calendar-day map and an eligible daily training target in the preceding five days. We evaluate every predictor on these same cases.}
\label{tab:operational-counts}
\setlength{\tabcolsep}{4pt}
\begin{tabular}{lrrrrrrr}\toprule
Fire & Starts & 1 day & 2 days & 3 days & 4 days & 5 days & Pairs\\\midrule
Bear & 12 & 11 & 11 & 10 & 9 & 8 & 49\\
Brattain & 8 & 7 & 7 & 6 & 5 & 5 & 30\\
Buck & 18 & 17 & 16 & 15 & 14 & 14 & 76\\
Chimney & 5 & 5 & 4 & 3 & 2 & 1 & 15\\
Ferguson & 19 & 19 & 18 & 17 & 16 & 15 & 85\\
Pier & 18 & 16 & 14 & 12 & 11 & 12 & 65\\
Total & 80 & 75 & 70 & 63 & 57 & 55 & 320\\
\bottomrule\end{tabular}
\end{table}

\begin{table}[H]\centering\small
\caption{Per-event daily-restart forecasts by fire. We use the same observed starts and target dates for every predictor and history policy. We average scores over starts within each fire and horizon. Bold values indicate the best result for each metric within the same fitting category and evaluation cases. Ties at the displayed precision share bold type.}
\label{tab:operational-per_event}
\setlength{\tabcolsep}{4pt}
\begin{tabular}{lllrrrrrr}\toprule
Fire & Predictor & History & 1-day MSE & IoU & 3-day MSE & IoU & 5-day MSE & IoU\\\midrule
Bear & Hybrid & Frozen & 0.0101 & 0.1222 & 0.0264 & 0.1303 & 0.0383 & 0.1798\\
 & Hybrid & Expanding & \textbf{0.0099} & 0.1467 & \textbf{0.0262} & 0.1545 & \textbf{0.0381} & 0.1914\\
 & Decoder & Frozen & 0.0122 & 0.1386 & 0.0282 & 0.1624 & 0.0424 & \textbf{0.2055}\\
 & Decoder & Expanding & 0.0115 & \textbf{0.1564} & 0.0266 & \textbf{0.1675} & 0.0394 & 0.1943\\
 & CA--CNN & Frozen & 0.0178 & 0.1238 & 0.0519 & 0.1640 & 0.0918 & 0.1407\\
Brattain & Hybrid & Frozen & 0.0040 & 0.0331 & \textbf{0.0049} & 0.0524 & 0.0061 & 0.0633\\
 & Hybrid & Expanding & \textbf{0.0039} & \textbf{0.1437} & \textbf{0.0049} & 0.1248 & \textbf{0.0060} & 0.1021\\
 & Decoder & Frozen & 0.0048 & 0.0776 & 0.0066 & 0.1309 & 0.0082 & 0.1485\\
 & Decoder & Expanding & 0.0046 & 0.0962 & 0.0062 & \textbf{0.1391} & 0.0077 & \textbf{0.2489}\\
 & CA--CNN & Frozen & 0.0109 & 0.0725 & 0.0709 & 0.0447 & 0.1657 & 0.0290\\
Buck & Hybrid & Frozen & \textbf{0.0049} & \textbf{0.2086} & 0.0139 & \textbf{0.2792} & 0.0241 & \textbf{0.2237}\\
 & Hybrid & Expanding & \textbf{0.0049} & 0.1912 & 0.0138 & 0.2389 & 0.0239 & 0.2163\\
 & Decoder & Frozen & 0.0050 & 0.2030 & 0.0136 & 0.2322 & 0.0237 & 0.1920\\
 & Decoder & Expanding & \textbf{0.0049} & 0.1907 & \textbf{0.0132} & 0.2321 & \textbf{0.0231} & 0.1989\\
 & CA--CNN & Frozen & 0.0092 & 0.0849 & 0.0252 & 0.1654 & 0.0381 & 0.2102\\
Chimney & Hybrid & Frozen & \textbf{0.0143} & 0.4351 & 0.0291 & 0.5001 & 0.0511 & 0.4933\\
 & Hybrid & Expanding & 0.0144 & 0.4288 & 0.0292 & 0.5037 & 0.0511 & 0.4933\\
 & Decoder & Frozen & 0.0239 & 0.3706 & 0.0386 & 0.5444 & 0.0504 & 0.5574\\
 & Decoder & Expanding & 0.0218 & 0.4048 & 0.0366 & 0.5410 & 0.0504 & 0.5574\\
 & CA--CNN & Frozen & 0.0216 & \textbf{0.4850} & \textbf{0.0211} & \textbf{0.7668} & \textbf{0.0258} & \textbf{0.7888}\\
Ferguson & Hybrid & Frozen & 0.0159 & \textbf{0.3696} & 0.0439 & 0.3766 & 0.0754 & 0.3892\\
 & Hybrid & Expanding & \textbf{0.0157} & 0.3617 & 0.0437 & 0.3957 & 0.0745 & 0.3998\\
 & Decoder & Frozen & 0.0175 & 0.3553 & \textbf{0.0435} & 0.3911 & 0.0704 & 0.3607\\
 & Decoder & Expanding & 0.0172 & 0.3507 & \textbf{0.0435} & 0.4039 & \textbf{0.0702} & 0.3946\\
 & CA--CNN & Frozen & 0.0376 & 0.3225 & 0.0763 & \textbf{0.4923} & 0.0725 & \textbf{0.6127}\\
Pier & Hybrid & Frozen & \textbf{0.0041} & \textbf{0.2477} & 0.0141 & 0.2997 & 0.0199 & 0.2821\\
 & Hybrid & Expanding & \textbf{0.0041} & 0.2153 & 0.0142 & 0.3049 & 0.0200 & 0.2890\\
 & Decoder & Frozen & 0.0042 & 0.2444 & \textbf{0.0135} & 0.2689 & \textbf{0.0185} & 0.2359\\
 & Decoder & Expanding & 0.0043 & 0.2176 & \textbf{0.0135} & 0.2903 & 0.0186 & 0.2736\\
 & CA--CNN & Frozen & 0.0067 & 0.2069 & 0.0198 & \textbf{0.3403} & 0.0287 & \textbf{0.3684}\\
\bottomrule\end{tabular}
\end{table}

\begin{table}[H]\centering\small
\caption{Pooled daily-restart forecasts by fire. We use the same observed starts and target dates for every predictor and history policy. We average scores over starts within each fire and horizon. Bold values indicate the best result for each metric within the same fitting category and evaluation cases. Ties at the displayed precision share bold type.}
\label{tab:operational-pooled}
\setlength{\tabcolsep}{4pt}
\begin{tabular}{lllrrrrrr}\toprule
Fire & Predictor & History & 1-day MSE & IoU & 3-day MSE & IoU & 5-day MSE & IoU\\\midrule
Bear & Hybrid & Frozen & \textbf{0.0096} & 0.1324 & 0.0247 & 0.1139 & 0.0347 & 0.1528\\
 & Hybrid & Expanding & \textbf{0.0096} & 0.1761 & \textbf{0.0246} & 0.1665 & \textbf{0.0345} & 0.1930\\
 & Decoder & Frozen & 0.0113 & 0.1391 & 0.0263 & 0.1541 & 0.0376 & 0.1817\\
 & Decoder & Expanding & 0.0112 & \textbf{0.1783} & 0.0260 & \textbf{0.1909} & 0.0369 & \textbf{0.2096}\\
 & CA--CNN & Frozen & 0.0194 & 0.1253 & 0.0548 & 0.1321 & 0.0923 & 0.1308\\
Brattain & Hybrid & Frozen & \textbf{0.0038} & 0.1500 & \textbf{0.0044} & 0.2255 & \textbf{0.0054} & 0.2326\\
 & Hybrid & Expanding & \textbf{0.0038} & 0.1733 & \textbf{0.0044} & 0.2130 & \textbf{0.0054} & 0.2132\\
 & Decoder & Frozen & 0.0040 & \textbf{0.1797} & 0.0048 & \textbf{0.2941} & 0.0060 & \textbf{0.3055}\\
 & Decoder & Expanding & 0.0040 & 0.1481 & 0.0048 & 0.2124 & 0.0060 & 0.2257\\
 & CA--CNN & Frozen & 0.0061 & 0.1221 & 0.0136 & 0.2079 & 0.0244 & 0.1937\\
Buck & Hybrid & Frozen & \textbf{0.0048} & 0.1738 & 0.0130 & 0.1732 & 0.0228 & 0.1228\\
 & Hybrid & Expanding & \textbf{0.0048} & 0.2002 & 0.0130 & 0.2528 & 0.0227 & \textbf{0.2689}\\
 & Decoder & Frozen & 0.0051 & \textbf{0.2098} & \textbf{0.0122} & 0.2427 & 0.0216 & 0.1797\\
 & Decoder & Expanding & 0.0051 & 0.1943 & \textbf{0.0122} & \textbf{0.2581} & \textbf{0.0214} & 0.2472\\
 & CA--CNN & Frozen & 0.0086 & 0.1350 & 0.0279 & 0.1599 & 0.0510 & 0.1602\\
Chimney & Hybrid & Frozen & \textbf{0.0140} & 0.4195 & 0.0301 & 0.4344 & 0.0545 & 0.4094\\
 & Hybrid & Expanding & \textbf{0.0140} & 0.4057 & 0.0301 & 0.4662 & 0.0545 & 0.4094\\
 & Decoder & Frozen & 0.0151 & 0.4146 & 0.0328 & 0.4254 & 0.0600 & 0.3898\\
 & Decoder & Expanding & 0.0150 & 0.3320 & 0.0327 & 0.4507 & 0.0600 & 0.3898\\
 & CA--CNN & Frozen & 0.0168 & \textbf{0.5221} & \textbf{0.0168} & \textbf{0.7654} & \textbf{0.0131} & \textbf{0.8279}\\
Ferguson & Hybrid & Frozen & \textbf{0.0157} & \textbf{0.3645} & 0.0426 & 0.3988 & 0.0726 & 0.4013\\
 & Hybrid & Expanding & \textbf{0.0157} & 0.3613 & 0.0426 & \textbf{0.4058} & 0.0725 & \textbf{0.4150}\\
 & Decoder & Frozen & 0.0165 & 0.3556 & \textbf{0.0422} & 0.3794 & 0.0698 & 0.3630\\
 & Decoder & Expanding & 0.0165 & 0.3531 & \textbf{0.0422} & 0.3912 & \textbf{0.0697} & 0.3798\\
 & CA--CNN & Frozen & 0.0241 & 0.2541 & 0.0579 & 0.3375 & 0.0864 & 0.4000\\
Pier & Hybrid & Frozen & \textbf{0.0041} & 0.2343 & 0.0138 & 0.3078 & 0.0194 & 0.2811\\
 & Hybrid & Expanding & \textbf{0.0041} & 0.2110 & 0.0138 & 0.3236 & 0.0194 & 0.3062\\
 & Decoder & Frozen & 0.0042 & \textbf{0.2433} & \textbf{0.0131} & 0.3017 & \textbf{0.0180} & 0.2593\\
 & Decoder & Expanding & 0.0043 & 0.2124 & \textbf{0.0131} & 0.3003 & \textbf{0.0180} & 0.2676\\
 & CA--CNN & Frozen & 0.0054 & 0.2269 & 0.0144 & \textbf{0.4066} & 0.0187 & \textbf{0.4214}\\
\bottomrule\end{tabular}
\end{table}

\section{Observed progression and binary predictions}
\label{app:maps}
\begin{figure}[p]\centering
\includegraphics[width=\linewidth,height=0.87\textheight,keepaspectratio]{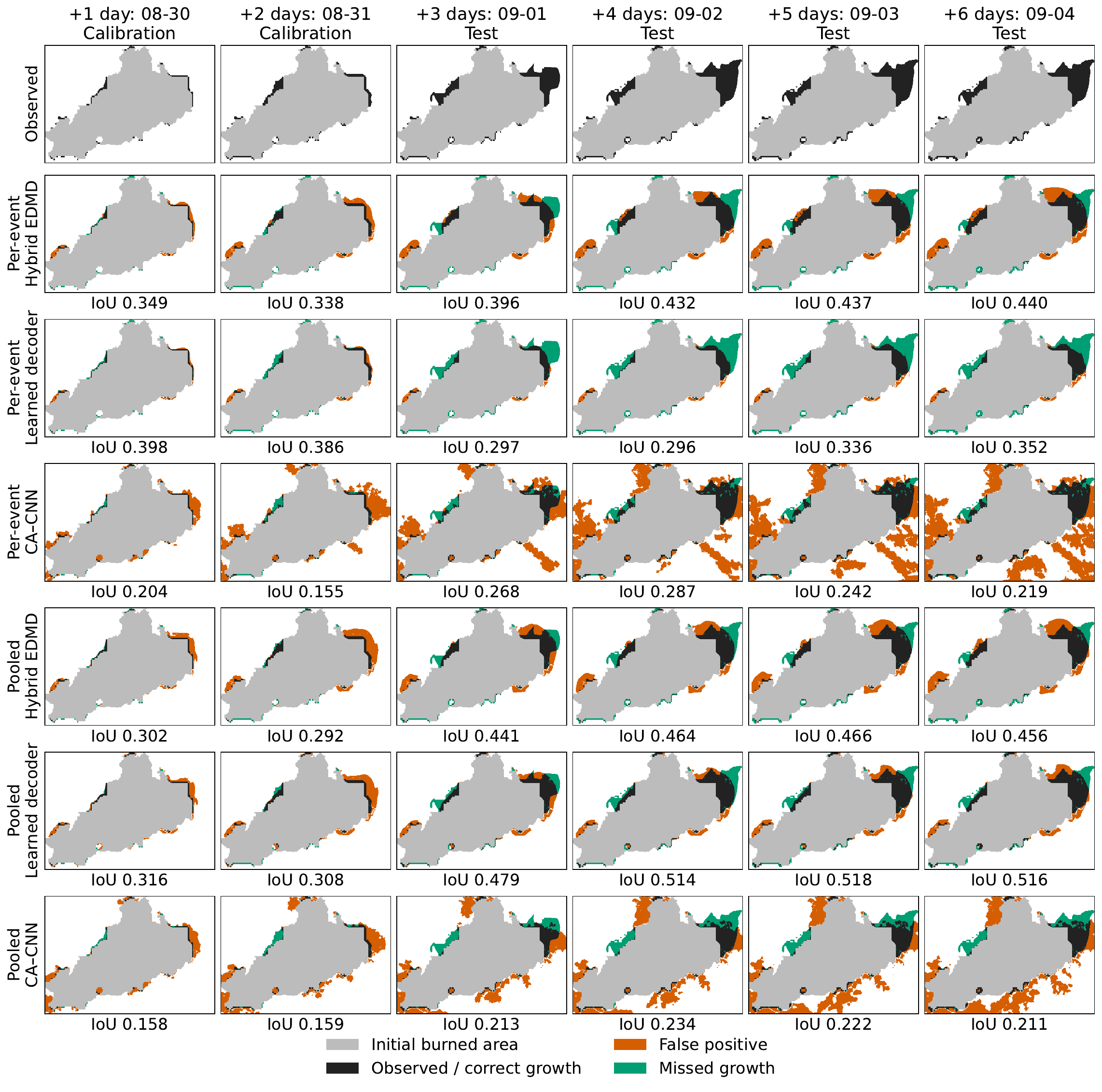}
\caption{Bear 2020. Continuous forecasts from the observed state at $d_{11}$. Columns show calendar dates, calibration or test status, and the target-date IoU outside the initial burned area.}\label{fig:map-bear_2020}
\end{figure}

\begin{figure}[p]\centering
\includegraphics[width=\linewidth,height=0.87\textheight,keepaspectratio]{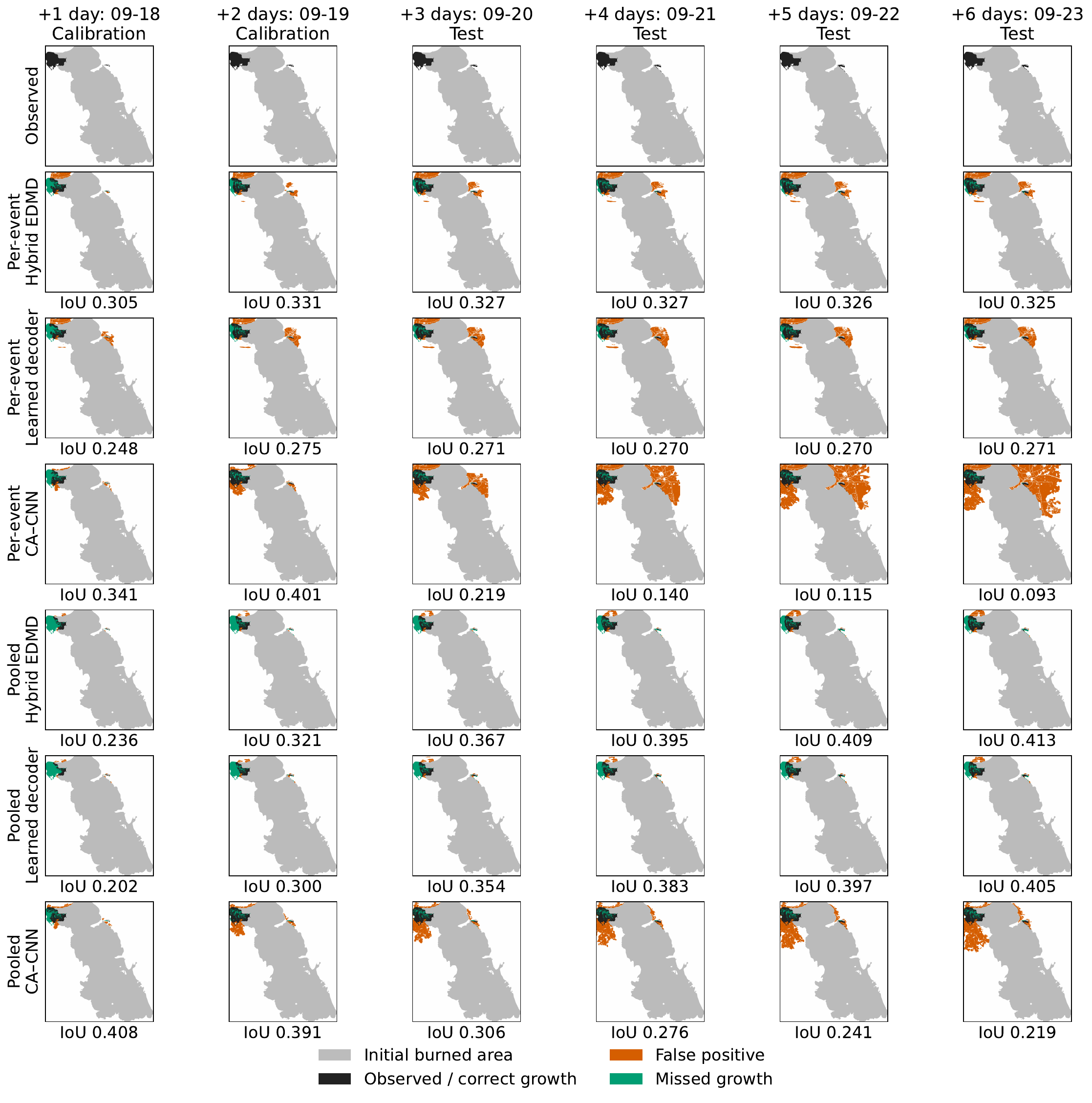}
\caption{Brattain 2020. Continuous forecasts from the observed state at $d_{11}$. Columns show calendar dates, calibration or test status, and the target-date IoU outside the initial burned area.}\label{fig:map-brattain_2020}
\end{figure}

\begin{figure}[p]\centering
\includegraphics[width=\linewidth,height=0.87\textheight,keepaspectratio]{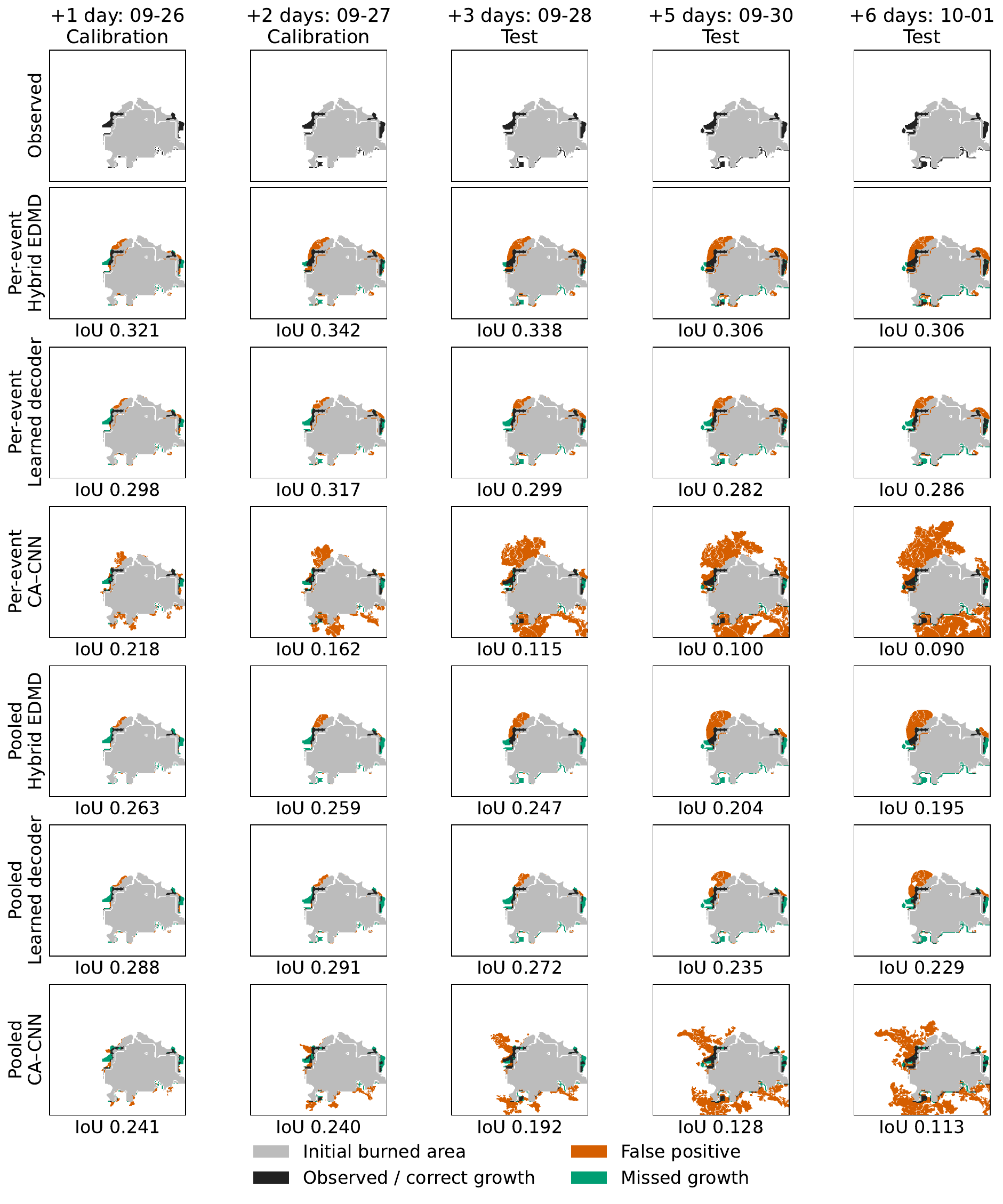}
\caption{Buck 2017. Continuous forecasts from the observed state at $d_{11}$. Columns show calendar dates, calibration or test status, and the target-date IoU outside the initial burned area.}\label{fig:map-buck_2017}
\end{figure}

\begin{figure}[p]\centering
\includegraphics[width=\linewidth,height=0.87\textheight,keepaspectratio]{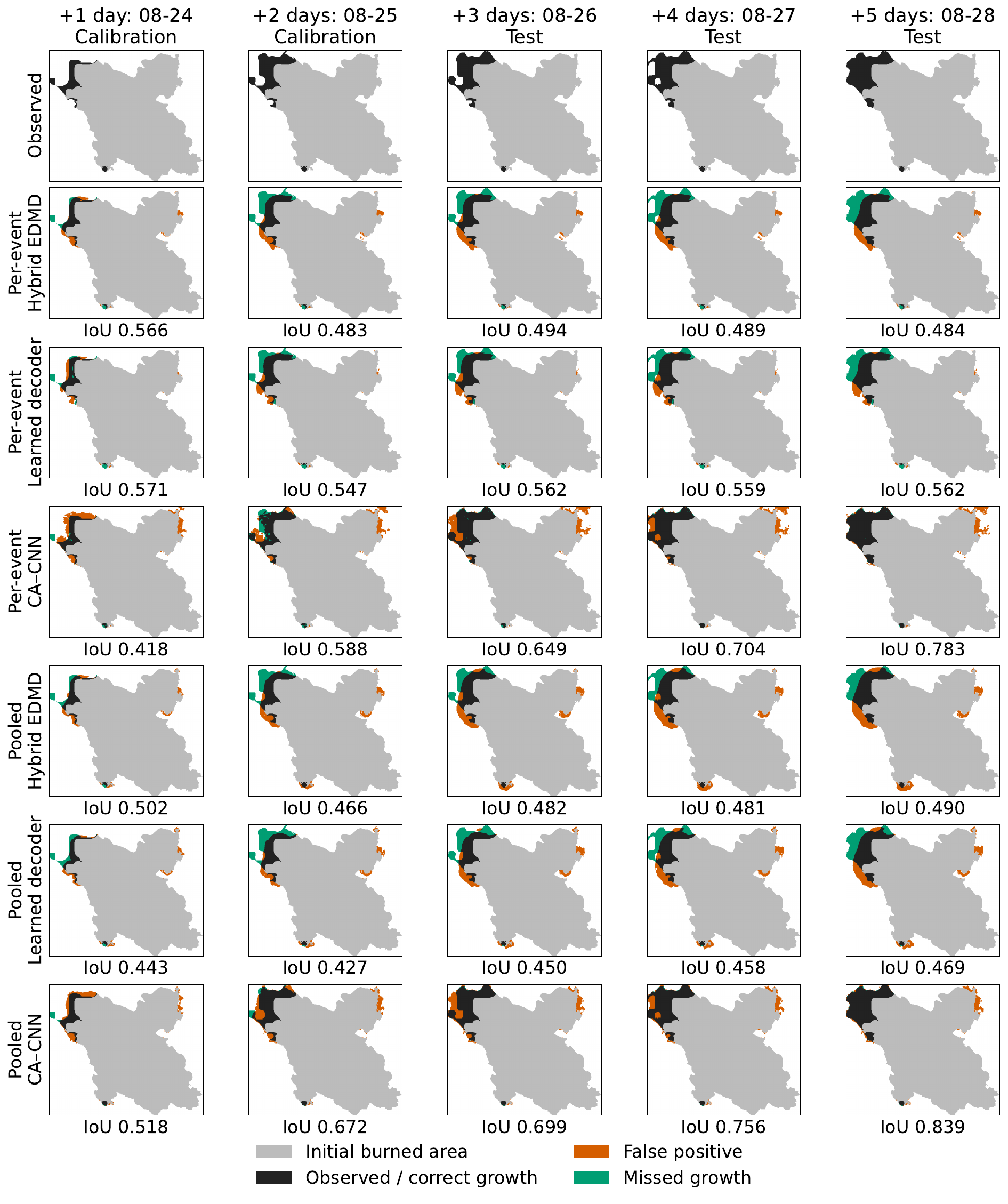}
\caption{Chimney 2016. Continuous forecasts from the observed state at $d_{11}$. Columns show calendar dates, calibration or test status, and the target-date IoU outside the initial burned area.}\label{fig:map-chimney_2016}
\end{figure}

\begin{figure}[p]\centering
\includegraphics[width=\linewidth,height=0.87\textheight,keepaspectratio]{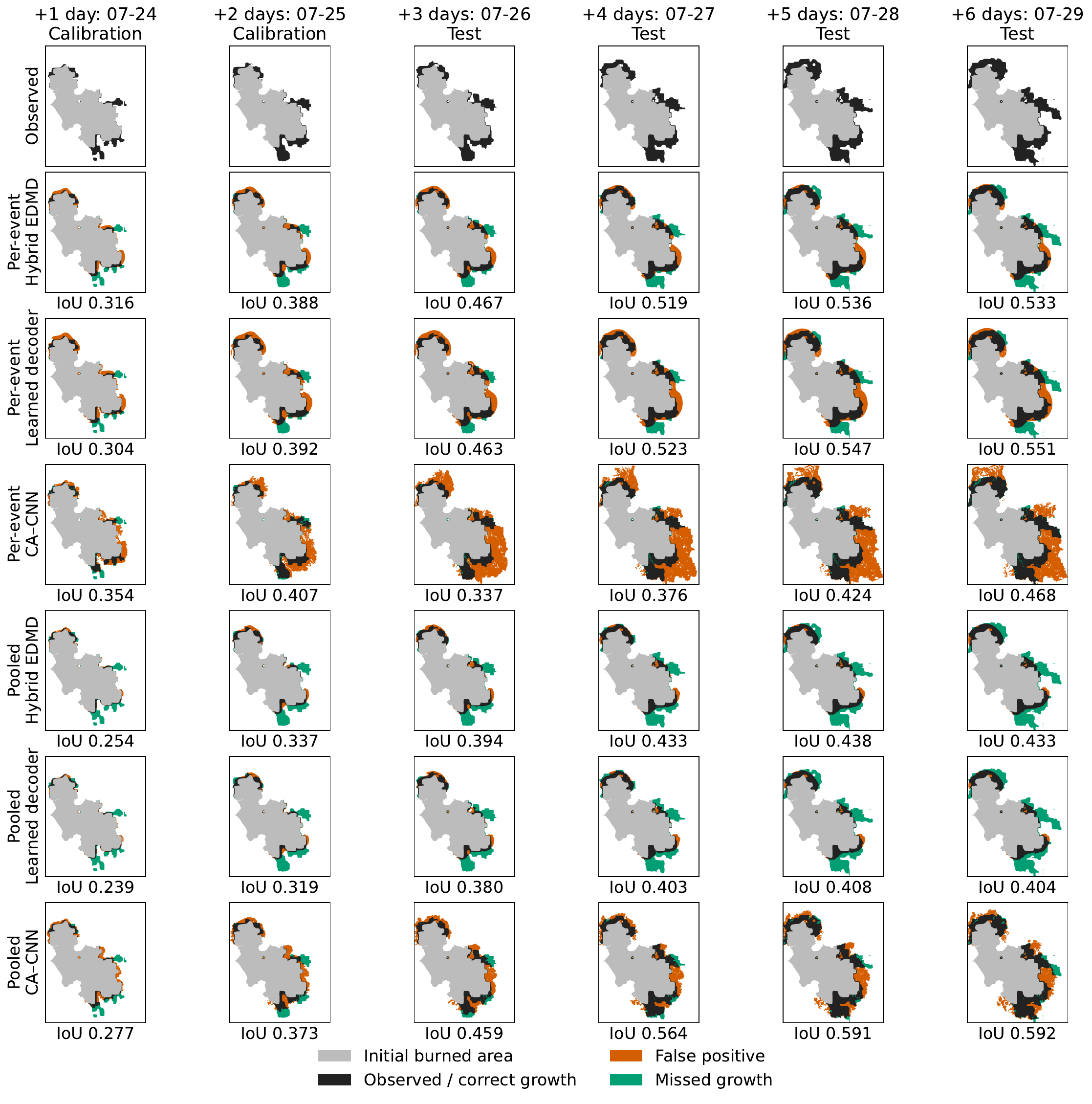}
\caption{Ferguson 2018. Continuous forecasts from the observed state at $d_{11}$. Columns show calendar dates, calibration or test status, and the target-date IoU outside the initial burned area.}\label{fig:map-ferguson_2018}
\end{figure}

\begin{figure}[p]\centering
\includegraphics[width=\linewidth,height=0.87\textheight,keepaspectratio]{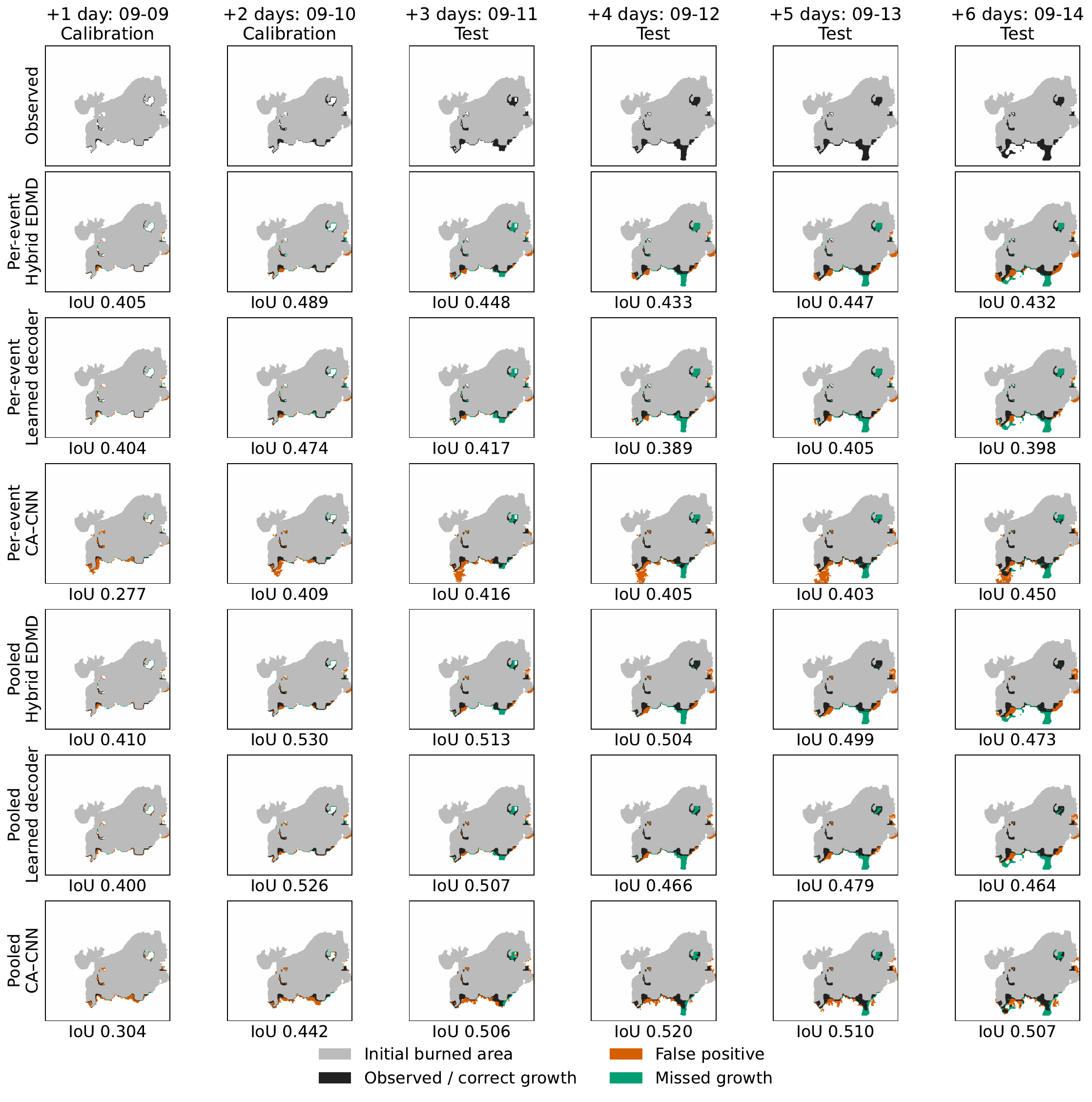}
\caption{Pier 2017. Continuous forecasts from the observed state at $d_{11}$. Columns show calendar dates, calibration or test status, and the target-date IoU outside the initial burned area.}\label{fig:map-pier_2017}
\end{figure}

Figures~\ref{fig:map-bear_2020} to~\ref{fig:map-pier_2017} show the six
fires on every observed date from day $t_0+1$ to day $t_0+6$, for both EDMD
predictors and CA--CNN, per event and pooled. All forecasts continue from
the observed state at $d_{11}$, and the first two columns are the
calibration days. Gray marks the initial burned area, excluded from
scoring. Black marks observed growth in the observation row and correct
predictions in the forecast rows, orange false positives, and green missed
growth. Each panel shows the full domain in raster orientation, with rows
increasing downward and columns to the right. Dates without a map are
omitted. Chimney has only five dates after the training period, so its
figure has fewer columns.

\clearpage

\clearpage

\clearpage

\clearpage

\clearpage

\clearpage

\end{document}